\documentclass[aps,prd,floats,floatfix,nofootinbib,preprintnumbers]{revtex4-1}
\usepackage{graphicx,amsmath}
\usepackage{amssymb}
\usepackage{amsfonts}
\usepackage{hyperref}
\usepackage[bottom]{footmisc}
\usepackage{subfigure}

\begin{document}

\preprint{MSUHEP-111013}

\title{Technipion Limits from LHC Higgs Searches }

\author{R.\ Sekhar Chivukula}
\email {sekhar@msu.edu}
\affiliation {Department of Physics and Astronomy,
Michigan State University,
East Lansing, MI 48824, USA}

\author{Pawin Ittisamai}
\email {ittisama@msu.edu}
\affiliation {Department of Physics and Astronomy,
Michigan State University,
East Lansing, MI 48824, USA}

\author{Jing Ren}
\email {jingren@pa.msu.edu}
\affiliation {Department of Physics and Astronomy,
Michigan State University,
East Lansing, MI 48824, USA}
\affiliation { Center for High Energy Physics and Institute of Modern Physics, 
Tsinghua University, 
Beijing 100084, China.}

\author{Elizabeth H.\ Simmons}
\email {esimmons@pa.msu.edu}
\affiliation {Department of Physics and Astronomy,
Michigan State University,
East Lansing, MI 48824, USA}

\date{\today}

\begin{abstract}
LHC searches for the standard model Higgs Boson in $\gamma\gamma$ or $\tau\tau$  decay modes
place strong constraints on the light top-pion state predicted in technicolor models that include colored technifermions.
Compared with the standard Higgs Boson, the top-pions have an enhanced production rate (largely because the technipion decay constant
is smaller than the weak scale) and also enhanced branching ratios into di-photon and di-tau final states (largely due to the suppression of
$WW$ decays of the technipions).  These factors combine to make the technipions more visible in both channels than a standard model Higgs would be.  Hence, the recent ATLAS and CMS searches for Higgs bosons exclude the presence of technipions with masses from 110 GeV to nearly $2 m_t$ in technicolor models that (a) include colored technifermions (b) feature topcolor dynamics and (c) have technicolor groups with three or more technicolors ($N_{TC} \geq 3$).   For certain models, the limits also apply out to higher technipion masses or down to the minimum number of technicolors  ($N_{TC} = 2$).   The limits may be softened somewhat in models where extended technicolor plays a significant role in producing the top quark's mass.  Additional LHC data on di-tau and di-photon final states will be extremely valuable in further exploring technicolor parameter space.

\end{abstract}

\maketitle

\section{Introduction}

Experiments now underway at the Large Hadron Collider are striving to discover the agent of electroweak symmetry breaking, thereby revealing the origin of the masses of the elementary particles.   Many of the searches are phrased in terms of placing constraints
on the properties of the scalar Higgs boson state predicted to exist in the standard model \cite{standard-model-a, standard-model-b, standard-model-c}.  In that theory, 
electroweak symmetry breaking occurs through the vacuum expectation value of a fundamental weak-doublet
scalar boson. Via the Higgs mechanism \cite{higgs-mechanism-a, higgs-mechanism-aa, higgs-mechanism-b, higgs-mechanism-c}, three of the scalar degrees of freedom of
this particle become the longitudinal states of the electroweak $W^\pm$ and $Z$ bosons and the last, the standard
model Higgs boson ($h_{SM}$), remains in the spectrum. Recently, both the ATLAS  and CMS collaborations at the CERN LHC have reported searches for the standard model Higgs in the two-photon \cite{Collaboration:2011ww, CMS-higgs-di-gamma}
 and $\tau^+ \tau^-$ \cite{ATLAS-higgs-di-tau-low, ATLAS-higgs-di-tau-high, CMS-higgs-di-tau} decay channels.  They have placed upper bounds on the cross-section times branching ratio ($\sigma \cdot B$)  in each channel over the approximate mass range 110 GeV $\leq m_{h} \leq$ 145 GeV, generally finding that $\sigma\cdot B$ cannot exceed the standard model prediction by more than a factor of a few.  In addition, ATLAS has independently constrained the production of a heavy neutral scalar SM Higgs boson with mass up to 600 GeV and decaying to $\tau^+ \tau^-$.  In this paper we apply these limits to the neutral ``technipion'' ($\Pi_T$) states predicted to exist in technicolor models that include colored technifermions.  Because both the technipion production rates and their branching fractions to $\gamma\gamma$ or $\tau\tau$ can greatly exceed the values for a standard model Higgs, the LHC results place strong constraints on technicolor models.  This strategy was first suggested as a possible for hadron supercolliders over fifteen years ago in Refs. \cite{Eichten:1984eu,Eichten:1986eq,Chivukula:1995dt}.
 
Technicolor \cite{Weinberg:1975gm, Weinberg:1979bn,Susskind:1978ms} is a dynamical theory of electroweak symmetry breaking in which a new strongly-coupled gauge group (technicolor) causes bilinears of the fermions carrying its gauge charge (technifermions) to acquire a non-zero vacuum expectation value.  If the technifermion bilinear carries appropriate weak and hypercharge values, the vacuum expectation value breaks the electroweak symmetry to its electromagnetic subgroup.  Fermion masses can then be produced dynamically if technicolor is incorporated into a larger ``extended technicolor''   \cite{Dimopoulos:1979es, Eichten:1979ah} framework coupling technifermions to the ordinary quarks and leptons.   Producing realistic values of fermion masses from extended technicolor (ETC) interactions without simultaneously generating large flavor-changing neutral currents (FCNC) is difficult; the best prospects are ``walking'' technicolor models where the presence of many technifermion flavors causes the technicolor gauge coupling to vary only slowly with energy scale \cite{Holdom:1981rm,Holdom:1984sk,Yamawaki:1985zg,Appelquist:1986an,Appelquist:1986tr,Appelquist:1987fc}.  Even in those models, it is difficult to generate the observed mass of the top quark from ETC interactions without producing unacceptably large weak isospin violation \cite{Chivukula:1988qr}; the best known solution is to generate most of the top quark's mass via new strong ``topcolor''   \cite{Hill:1991(a)t} dynamics, without a large contribution from ETC \cite{Hill:1994hp}.  

As we review below (see also \cite{Hill:2002ap}), many technicolor models, including those with walking and topcolor dynamics, feature technipion states, pseudo-scalar bosons that are remnants of electroweak symmetry breaking in models with more than one weak doublet of technifermions.   Production of light technipion states at lepton colliders has been studied by a variety of authors \cite{Manohar:1990eg, Randall:1991gp, Rupak:1995kg, Lubicz:1995xi, Casalbuoni:1998fs, Lynch:2000hi}; the most comprehensive analysis \cite{Lynch:2000hi} used LEP I and LEP II data to constrain the anomalous couplings of technipions to neutral electroweak gauge bosons and derived limits on the size of the technicolor gauge group and the number of technifermion doublets in various representative technicolor models.  Subsequently,  the authors of \cite{Belyaev:2005ct} considered technipion phenomenology at hadron colliders; they demonstrated both that technipions can be produced at a greater rate than the standard model Higgs, because the technipion decay constant is smaller than the electroweak scale, and also that the technipions can also have higher branching fractions to $\gamma\gamma$ or $\tau\tau$ final states.  As a result, the technipions are predicted to produce larger signals in these two channels at LHC than the $h_{SM}$ would \cite{Belyaev:2005ct}.   

In this work, we show that the ATLAS \cite{Collaboration:2011ww, ATLAS-higgs-di-tau-low, ATLAS-higgs-di-tau-high} and CMS \cite{CMS-higgs-di-gamma, CMS-higgs-di-tau} searches for the standard model Higgs exclude, at 95\% CL, technipions of masses from 110 GeV to nearly $2 m_t$ in technicolor models that (a) include colored technifermions (b) feature topcolor dynamics and (c) have technicolor groups with three or more technicolors ($N_{TC} \geq 3$).   For certain models of this kind, the limits also apply out to higher technipion masses or down to the minimum number of technicolors  ($N_{TC} = 2$).  We also show how the limits may be modified in models in which extended technicolor plays a significant role in producing the mass of the top quark; in some cases, this makes little difference, while in other cases the limit is softened somewhat.  Overall, we find that the ATLAS and CMS significantly constrain technicolor models.  Moreover, as the LHC collaborations collect additional data on these di-tau and di-photon final states and extend the di-photon analyses to higher mass ranges, they should be able to quickly expand their reach in technicolor parameter space.

\section{Technicolor and Technipions}
\label{asec:technicp}

 Dynamical theories of electroweak symmetry breaking embody the possibility that the scalar states involved in electroweak symmetry breaking could be manifestly composite at scales not much above the weak scale $v  \approx 246$ GeV.
In technicolor theories \cite{Susskind:1978ms,Weinberg:1975gm, Weinberg:1979bn}, a new asymptotically free strong gauge interaction  
breaks the chiral symmetries of massless fermions $T$ at
a scale $\Lambda \sim 1$ TeV.  If the fermions carry appropriate
electroweak quantum numbers (e.g. left-hand (LH) weak doublets and right-hand (RH) weak
singlets), the resulting condensate $\langle \bar T_L T_R \rangle \neq
0$ breaks the electroweak symmetry correctly to its electromagnetic subgroup.  Three of the Nambu-Goldstone Bosons
 of the chiral symmetry breaking become the longitudinal modes of the $W$ and $Z$, making those gauge bosons massive. The hierarchy and triviality problems plaguing the standard model are absent:  the logarithmic running of the
strong gauge coupling renders the low value of the electroweak scale natural, while the absence of fundamental scalars obviates concerns about triviality.  

In so-called minimal technicolor models, there are no composite scalars left in the spectrum.  However, many dynamical symmetry-breaking models include more than the minimal two flavors of technifermions needed to break the electroweak symmetry.  In that case, there will exist light pseudo Nambu-Goldstone bosons known as technipions, which could potentially be accessible to a standard Higgs search. Technipions that are bound states of colored technifermions can be produced through quark or gluon scattering at a hadron collider, like the LHC, through the diagrams in Figure 1.   In the models with topcolor dynamics, where ETC interactions (represented by the shaded circle) contribute no more than a few GeV to the mass of any quark, there is only a small ETC-mediated coupling between the technipion and ordinary quarks in diagrams 1(b) and 1(c).  Combining that information with the large size of the gluon PDF at the LHC and the $N_{TC}$ enhancement factor in the techniquark loop at left, we expect that the diagram in Figure 1(a) will dominate technipion production in these theories, which we study here and in Section \ref{asec:topcol}.  Technipions in models without strong top dynamics could, in contrast, have a large top-technipion coupling, making diagram 1(c) potentially important; we will consider that scenario in Section \ref{asec:topetc}.  Technipions that are bound states of non-colored technifermions would be produced at hadron colliders only through diagrams 1(b) and 1(c),  which would generally yield a significantly lower production rate; we comment on these models in the discussion (Section \ref{axax}).

	\begin{figure}[t,b]
	\centering
	\subfigure[\  gluon-gluon fusion through techniquark loop]{
		\includegraphics[scale=0.25]{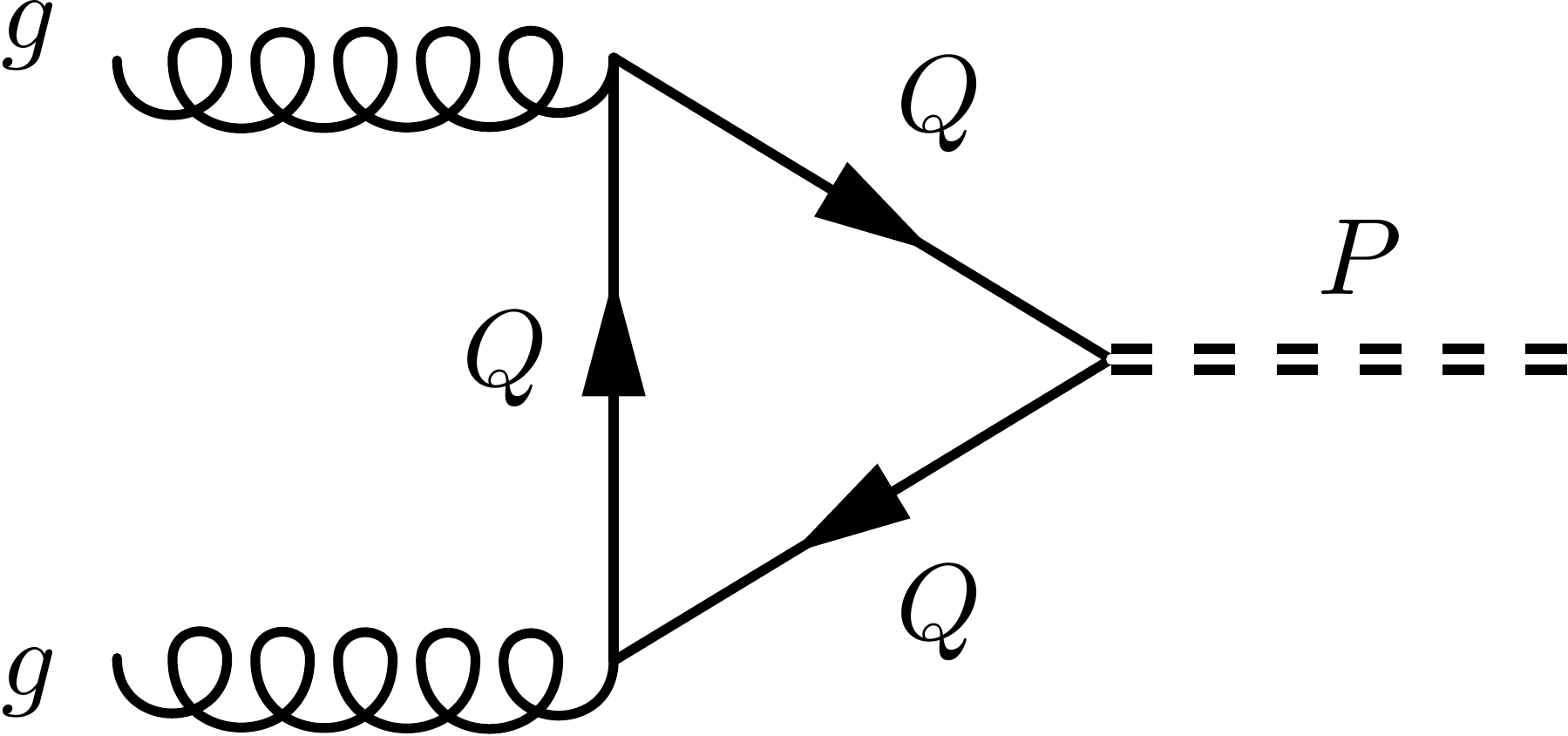}
	} \qquad\qquad
	\subfigure[\ $b\bar{b}$ annihilation]{
		\includegraphics[scale=0.25]{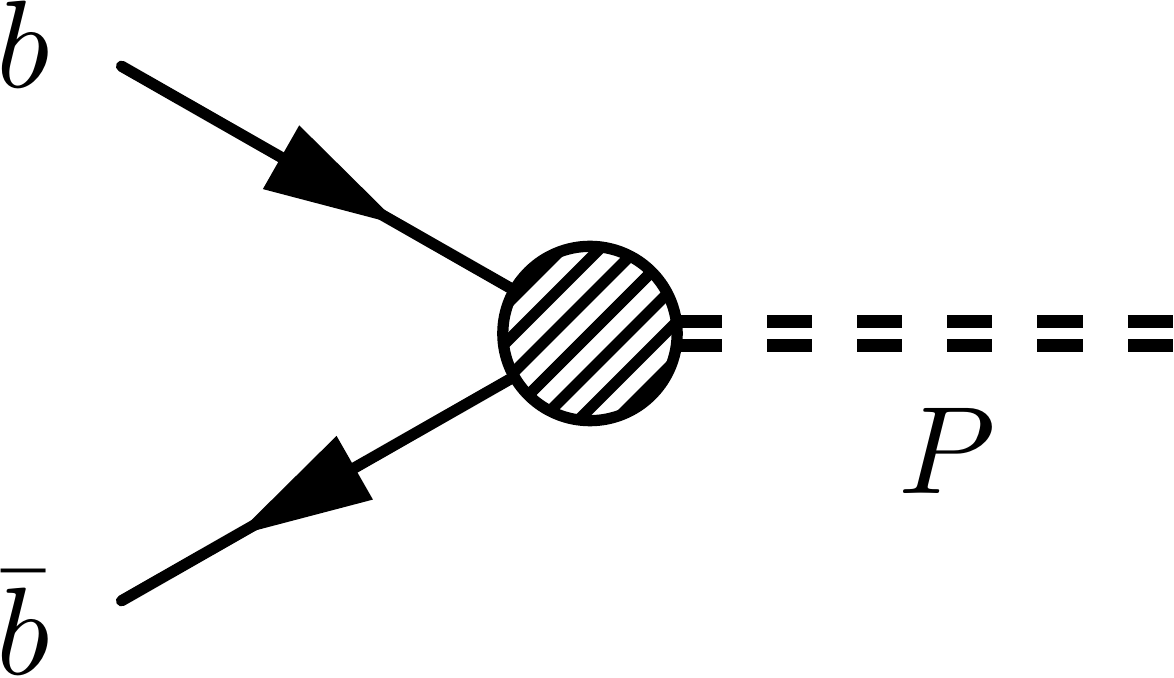}
	} \qquad\qquad
	\subfigure[\ gluon-gluon fusion through top quark loop]{
		\includegraphics[scale=0.25]{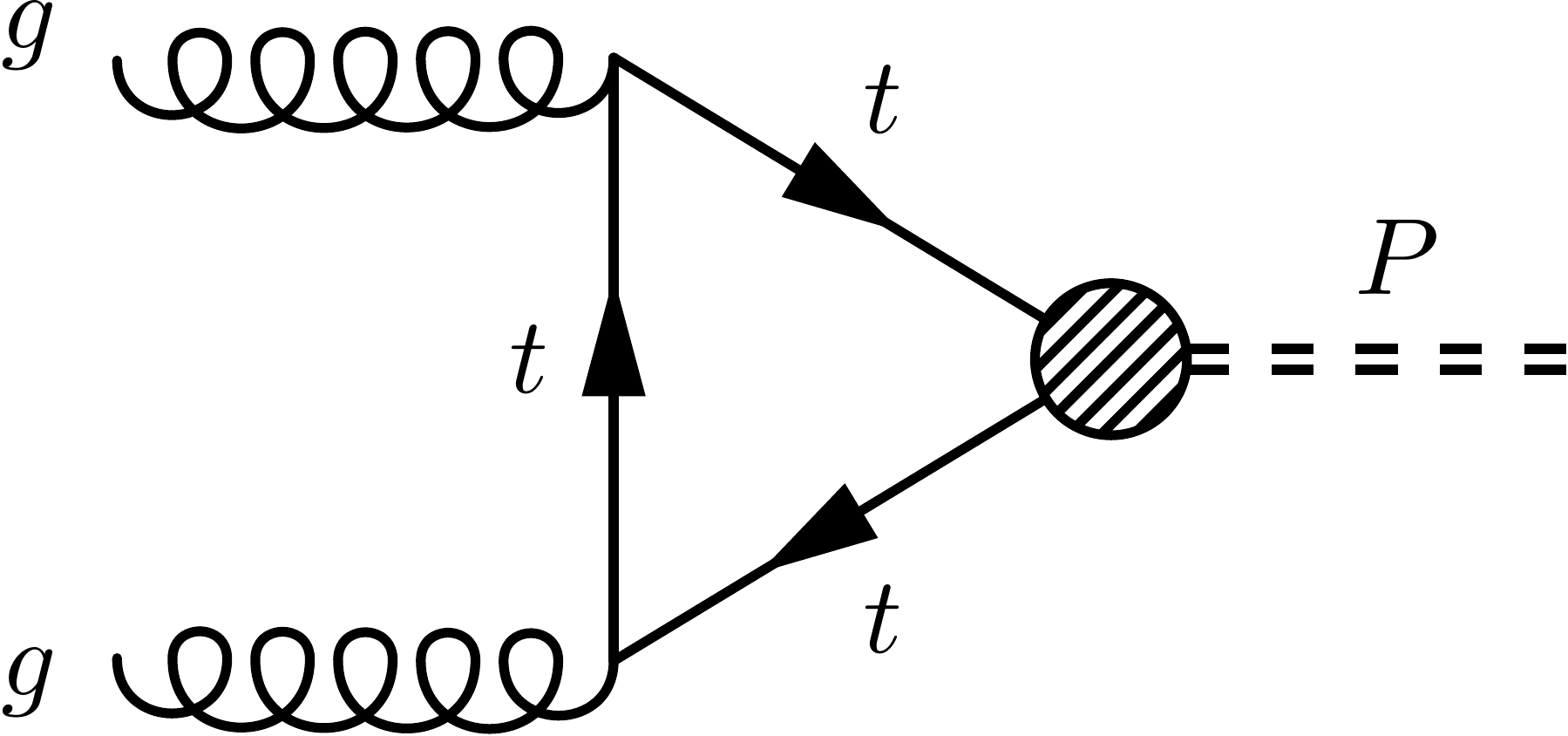}
	}
	\caption{Feynman diagrams for single technipion production at LHC.  The shaded circle in diagrams (b) and (c) represents an ETC coupling between the ordinary quarks and techniquarks.}
	\label{fig:feynman-diagrams}
	\end{figure}

No single technicolor model has been singled out as a benchmark; rather, different classes of models have been proposed to address the challenges of dynamically generating mass while complying with precision electroweak and flavor constraints.   We will study the general constraints that the current LHC data can place a variety of theories with colored technifermions and light technipions.  Following \cite{Belyaev:2005ct, Lynch:2000hi}, the specific models we examine are: 1) the original one-family model of Farhi and Susskind \cite{Farhi:1980xs} with a full family of techniquarks and technileptons, 2) a variant on the one-family model \cite{Casalbuoni:1998fs} in which the lightest technipion contains only down-type technifermions and is significantly lighter than the other pseudo Nambu-Goldstone
bosons,  3) a multiscale walking technicolor model \cite{Lane:1991qh} designed to reduce flavor-changing neutral currents,
4) a low-scale technciolor model (the Technicolor Straw Man -- TCSM -- model) \cite{Lane:1999uh} with many weak doublets of technifermions and 5) a one-family models with weak-isotriplet technifermions \cite{Manohar:1990eg}.  Properties of the lightest electrically-neutral technipion in each model that couples to gluons (and can therefore be readily produced at LHC) are shown in Table \ref{tab:technipions}.  For completeness, we show the name and technifermion content of each state in the notation of the original paper proposing its existence;  while each paper has its own conventions, all technifermion names including ``Q" or ``D" refer to color-triplets (a.k.a. techniquarks) while those including ``L" or ``E" refer to color-singlets (a.k.a. technileptons).\footnote{Note that the LR multiscale model  \protect\cite{Lane:1991qh} incorporates six technileptons, which
we denote $L_{\ell}$.}  In the TCSM low-scale model, the second-lightest technipion is the state relevant for our study (the lightest, being composed of technileptons, lacks an  anomalous coupling to gluons); in the other models the lightest technipion is the relevant one.  For simplicity the lightest relevant neutral technipion of each model will be generically denoted $P$. Furthermore, we will assume that the lightest technipion state is significantly lighter than other neutral (pseudo)scalar technipions in the spectrum, in order to facilitate the comparison to the standard model Higgs boson.\footnote{ The detailed spectrum of any technicolor model depends on multiple factors, particularly the parameters describing the  ``extended technicolor" \cite{Dimopoulos:1979es,Eichten:1979ah} interaction that transmits electroweak symmetry breaking to the ordinary quarks and leptons.  Models in which several light neutral PNGBs are nearly degenerate could produce even larger signals than those discussed here.}

Single production of a technipion can occur through the axial-vector anomaly which
couples the technipion to pairs of gauge bosons. For an $SU(N_{TC})$ technicolor group
with technipion decay constant $F_P$, the anomalous coupling between the technipion and
a pair of gauge bosons is given, in direct analogy with the coupling of a QCD pion to
photons,\footnote{Note that the normalization used here is identical to that in \cite{Belyaev:2005ct} and differs
from that used in \cite{Lynch:2000hi} by a factor of 4.}
 by \cite{Dimopoulos:1980yf,Ellis:1980hz,Holdom:1981(b)g}
\begin{equation}
N_{TC} {\cal A}_{V_1 V_2} {\frac{g_1 g_2}{8 \pi^2 F_P}} \epsilon_{\mu\nu\lambda\sigma}
k_1^\mu k_2^\nu \epsilon_1^\lambda \epsilon_2^\sigma
\label{eq:anom}
\end{equation}
where 
\begin{equation}
{\cal A}_{V_1 V_2} \equiv Tr \left[ T^a ( T_1 T_2 + T_2 T_1)_L + T^a (T_1 T_2 + T_2 T_1)_R \right]
\end{equation}
is the anomaly factor, $T^a$ is the generator of the axial vector current associated with the techipion, subscripts $L$ and $R$ denote the left- and right-handed technifermion components of the technipion, the $T_i$ and $g_i$ are the generators and couplings associated with gauge bosons $V_i$, and the $k_i$ and $\epsilon_i$ are the four-momenta and polarizations of the gauge
bosons. The value of the anomaly factor ${\cal A}_{gg}$ for the lightest PNGB of each model that is capable of coupling to gluons appears in Table 
\ref{tab:technipions}, along with the anomaly factor ${\cal A}_{\gamma\gamma}$ coupling the PNGB to photons.  Also shown in the table is the value of the technipion decay constant, $F_P$ for each model.\footnote{ In the multi-scale model [model 3], various technicondensates form at different scales; we set $F_P^{(3)}$ = $ \frac{v}{4}$ in keeping with \cite{Lane:1991qh}  and to ensure that the technipion mass will be in the range to which the standard Higgs searches are sensitive. }

Examining the technipion wavefunctions in Table \ref{tab:technipions} we note that the PNGB's do {\it not} decay to $W$ boson pairs, since the $W^+W^-$ analog of Figure 1(a)  vanishes due to a cancellation between techniquarks and technileptons.  The corresponding $ZZ$ diagrams will not vanish but, again due to a cancellation between techniquarks and technileptons, will instead yield small couplings for the technipion to
$ZZ$ (and $Z\gamma$) proportional to the technifermion hypercharge couplings \cite{Lynch:2000hi}. The small coupling and phase space suppression yield much smaller branching ratios for the PNGB's to decay
to $ZZ$ or $Z\gamma$, and hence these modes are irrelevant to our limits.

	\renewcommand{\arraystretch}{1.2}
	\begin{table}[!bt]
	\caption{Properties of the lightest relevant PNGB (technipion)  in representative technicolor models with colored technifermions.  In each case, we show the name and technifermion content of the state (in the notation of the original paper), the ratio of the weak scale to the technipion decay constant, the anomaly factors for the two-gluon and two-photon couplings of the technipion, and the technipion's  couplings to leptons and quarks.  The symbols``Q" or ``D" refer to color-triplets (a.k.a. techniquarks) while those including ``L" or ``E" refer to color-singlets (a.k.a. technileptons).  The multiscale model incorporates six technieptons, which we denote by $L_{\ell}$. For the TCSM low-scale model, $N_D$ refers to the number of weak-doublet technifermions contributing to electroweak symmetry breaking; this varies with the size of the technicolor group.    The parameter $y$ in the isotriplet model is the hypercharge assigned to the technifermions.}
\begin{center}
\begin{tabular}{c||c|c|c|c|c|c|c}
\hline\hline
TC models & \multicolumn{2}{c|}{PNGB and content} & $v/F_P$ & $A_{gg}$ & $A_{\gamma\gamma}$ & $\lambda_l$ & $\lambda_f$\\
\hline\hline
FS one family\cite{Farhi:1980xs} & $P^1$ & $\frac{1}{4\sqrt{3}}(3\bar{L}\gamma_5L-\bar{Q}\gamma_5Q)$ & 2 & $-\frac{1}{\sqrt{3}}$ & $\frac{4}{3\sqrt{3}}$ & 1 & 1 \\
\hline
Variant one family\cite{Casalbuoni:1998fs} & $P^0$ & $\frac{1}{2\sqrt{6}}(3\bar{E}\gamma_5E-\bar{D}\gamma_5D)$ & 1 & $-\frac{1}{\sqrt{6}}$ & $\frac{16}{3\sqrt{6}}$ & $\sqrt{6}$ & $\sqrt\frac{2}{3}$\\
\hline
LR multiscale\cite{Lane:1991qh} & $P^0$ & $\frac{1}{6\sqrt{2}}(\bar{L}_{\ell}\gamma_5L_{\ell}-2\bar{Q}\gamma_5Q)$ & 4 & $-\frac{2\sqrt{2}}{3}$ & $\frac{8\sqrt{2}}{9}$ & 1 & 1 \\
\hline
TCSM low scale\cite{Lane:1999uh} & $\pi^{0'}_T$ & $\frac{1}{4\sqrt{3}}(3\bar{L}\gamma_5L-\bar{Q}\gamma_5Q)$ & $\sqrt{N_D}$ & $-\frac{1}{\sqrt{3}}$ & $\frac{100}{27\sqrt{3}}$ & 1 & 1 \\
\hline
MR Isotriplet \cite{Manohar:1990eg} & $P^1$ & $\frac{1}{6\sqrt{2}}(3\bar{L}\gamma_5L-\bar{Q}\gamma_5Q)$ & 4 & $-\frac{1}{\sqrt{2}}$ & $24\sqrt{2}y^2$ & 1 & 1  \\
\hline\hline
\end{tabular}
\end{center}
	\label{tab:technipions}
	\end{table}

The rate of single technipion production via glue-glue fusion and a techniquark loop (Figure 1(a)) is proportional to the technipion's decay width to gluons through that same techniquark loop
\begin {equation}
{\Gamma (P \rightarrow gg)} = { \frac{m_{P}^3}{ 8 \pi}}   \left(\frac {\alpha_s N_{TC}{\cal A}_{gg}}{2 \pi F_P} \right)^2\ .
\label{eq:techni-glu}
\end {equation}
 In the  SM, the equivalent expression (for Higgs decay through a top quark loop) looks like \cite{Gunion:1989we} 
\begin {equation}
{\Gamma (h_{SM} \rightarrow gg)} ={\frac{m_{h}^3}{8 \pi}}  \left(\frac{ \alpha_s}{3 \pi v}\right)^2  
\left[\frac{3\tau}{2}(1 + (1-\tau) f(\tau))\right]^2~,
\label{eq:higgs-tau}
\end {equation}
where $\tau \equiv (4 m_t^2 / m_h^2)$ and
\begin{equation}
f(\tau) = 
\begin{cases} \left[ \sin^{-1} (\tau^{-{\frac12}}) \right]^2& \text{if $\tau \geq 1$}
\\
-\frac14 \left[ \log \left( \frac{1 + \sqrt{1-\tau}}{1 - \sqrt{1-\tau}}\right) - i \pi \right]^2&\text{if $\tau < 1$.}
\end{cases}
\label{eq:fftau}
\end{equation}
so that the expression in square brackets in Eq. (\ref{eq:higgs-tau}) approaches 1 in the limit where the top quark is heavy ($\tau >> 1$).  Therefore, the rate at which $P$ is produced from $gg$ fusion exceeds that for a standard Higgs of the same mass by a factor
\begin {equation}
\kappa_{gg\ prod} = \frac{ \Gamma (P \rightarrow gg)}{ \Gamma (h_{SM} \rightarrow gg)} = 
 \frac{9}{4} N_{TC}^2 {\cal A}_{gg}^2 \frac{v^2}{F_P^2} 
\left[\frac{3\tau}{2}(1 + (1-\tau) f(\tau))\right]^{-2}
\label{eq:kappagg}
\end {equation}
 where, again, the factor in square brackets is 1 for scalars much lighter than $2m_t$.  A large technicolor group and a small technipion decay constant can produce a significant enhancement factor.

Technipions can also be produced at hadron colliders via $b\bar{b}$ annihilation (as in Figure 1(b)), because the 
 ETC interactions coupling quarks to techniquarks afford the technipion a decay mode into fermion/anti-fermion pairs.  
The rate is proportional to the technipion decay width into fermions:
\begin {equation}
{ \Gamma (P \rightarrow f \overline{f})} = {\frac{N_C\, \lambda^2_f\, m^2_f\, m_P}{8 \pi\, F^2_P}}\,
\left(1 -  \frac{4m_f^2}{m_P^2}\right)^{\frac{s}{2}}
\label{eq:phasesp}
\end {equation}
where  $N_C$ is 3 for quarks and 1 for leptons.  The phase space exponent,
$s$, is 3 for scalars and 1 for pseudoscalars; the lightest PNGB in our technicolor models is a
pseudoscalar.  For the technipion masses considered here, the value of
the phase space factor in (\ref{eq:phasesp}) is so close to one that the
value of $s$ makes no practical difference.   The factors $\lambda_f$ are 
non-standard Yukawa couplings distinguishing leptons from quarks.  The variant one-family model has $\lambda_{quark} = \sqrt{\frac{2}{3}}$ and $\lambda_{lepton} =
\sqrt{6}$; the multiscale model also includes a similar factor, but with average
value 1; $\lambda_f=1$ in the other models.   For comparison, the decay width of the SM Higgs into b-quarks is:
\begin {equation}
{ \Gamma (h_{SM} \rightarrow b \overline{b})} = {\frac{3\,m^2_b\,m_h}{8\pi\,v^2}}
\left(1 - \frac {4m_b^2}{m_h^2}\right)^{\frac{3}{2}}
\end {equation}
 Thus, the rate at which $P$ is produced from $b\bar{b}$ annihilation exceeds that for a standard Higgs of the same mass by 
\begin {equation}
\kappa_{bb\ prod} = \frac{ \Gamma (P \rightarrow b \overline{b})}
{ \Gamma (h_{SM} \rightarrow b \overline{b})} = {\frac{\lambda^2_b\, v^2}{ F^2_P}} 
\left(1 - \frac{4m_b^2}{m_h^2}\right)^{\frac{s-3}{2}}
\label{eq:kappabb}
\end {equation}
The enhancement is smaller than that in Eq. (\ref{eq:kappagg}) because there is no loop-derived factor of $N_{TC}$.

\begin{table}[!tb]
\label{tab:BR130}
\caption{Branching ratios for phenomenologically important modes (in percent) for technipions of mass 130 GeV for $N_{TC}=2,4$ and for a standard model Higgs  \protect\cite{Dittmaier:2011ti} of the same mass.}
\begin{center}
\begin{tabular}{|c|| c|c|| c|c|| c|c|| c|c|| c|c| ||c|}
\hline
  &\multicolumn{2}{c||}{One} &\multicolumn{2}{c||}{Variant} &\multicolumn{2}{c||}{Multiscale} &\multicolumn{2}{c||}{TCSM} &\multicolumn{2}{c|||}{Isotriplet} &\multicolumn{1}{c|}{}\\
  Decay &\multicolumn{2}{c||}{Family} &\multicolumn{2}{c||}{one family} &\multicolumn{2}{c||}{} &\multicolumn{2}{c||}{low-scale} &\multicolumn{2}{c|||}{} &\multicolumn{1}{c|}{SM}\\
\cline{2-11}
Channel 				 &$N_{TC}$	&$N_{TC}$ 		&$N_{TC}$	&$N_{TC}$		&$N_{TC}$	&$N_{TC}$		&$N_{TC}$	&$N_{TC}$	&$N_{TC}$	&$N_{TC}$ 		& Higgs\\
					&=2 & =4	 	&=2 & =4	 	&=2 & =4	 	&=2 & =4	 		&=2 & =4	 			&  \\
\hline\hline
$b\bar{b}$	 		& 77 & 56		&61 & 50 	 	& 64& 36 		& 77 & 56 		 	&60& 31 			& 49\\
$c\bar{c}$	 		& 7 & 5.1		& 0 & 0	 		& 5.8 & 3.2		& 7 & 5.1	 		&5.4& 2.8			& 2.3\\
$\tau^+\tau^-$	 	& 4.5 & 3.3		&32	& 26	 	& 3.8& 2.1		& 4.5 & 3.3		 	&3.5& 1.8 			& 5.5 \\
$gg$				& 12 & 35		& 7& 23	 		& 26 & 59 		& 12 & 35  			&14& 29				& 7.9\\
$\gamma\gamma$	& 0.011 & 0.033	&0.11 & 0.35	& 0.025& 0.056	& 0.088 & 0.26	 	&17& 36				& 0.23\\
$W^+W^-$			& 0 & 0			& 0 & 0	 		& 0 & 0			& 0 & 0	 			&0 & 0 				& 31 \\
\hline
\end{tabular}
\end{center}
\end{table}

\begin{table}[!bt]
\label{tab:BR350}
\caption{Branching ratios for phenomenologically important modes (in percent) for technipions of mass 350 GeV for $N_{TC}=2,4$ and for a standard model Higgs  \protect\cite{Dittmaier:2011ti} of the same mass.}
\begin{center}
\begin{tabular}{|c|| c|c|| c|c|| c|c|| c|c|| c|c| ||c|}
\hline
  &\multicolumn{2}{c||}{One} &\multicolumn{2}{c||}{Variant} &\multicolumn{2}{c||}{Multiscale} &\multicolumn{2}{c||}{TCSM} &\multicolumn{2}{c|||}{Isotriplet} &\multicolumn{1}{c|}{}\\
  Decay &\multicolumn{2}{c||}{Family} &\multicolumn{2}{c||}{one family} &\multicolumn{2}{c||}{} &\multicolumn{2}{c||}{low-scale} &\multicolumn{2}{c|||}{} &\multicolumn{1}{c|}{SM}\\
\cline{2-11}
Channel 				 &$N_{TC}$	&$N_{TC}$ 		&$N_{TC}$	&$N_{TC}$		&$N_{TC}$	&$N_{TC}$		&$N_{TC}$	&$N_{TC}$	&$N_{TC}$	&$N_{TC}$ 		& Higgs\\
					&=2 & =4	 	&=2 & =4	 	&=2 & =4	 	&=2 & =4	 		&=2 & =4	 			&  \\
\hline\hline
$b\bar{b}$	 		& 44 & 18		&42 & 20 	 	& 24 & 7.7 		& 44 & 18 		 	&20& 6.2 			& 0.036\\
$c\bar{c}$	 		& 4 & 1.6		& 0 & 0	 		& 2.2 & 0.69		& 4 & 1.6	 		&1.8& 0.56			& 0.0017\\
$\tau^+\tau^-$	 	& 2.6 & 1		&22	& 11	 	& 1.4 & 0.45		& 2.6 & 1		 	&1.2& 0.36 			& 0.0048\\
$gg$				& 49 & 79		& 35& 68 		& 72 & 91 		& 49 & 79  			&34& 41				& 0.085\\
$\gamma\gamma$	& 0.047 & 0.076	&0.54 & 1		& 0.069& 0.087	& 0.36 & 0.58 		&42& 51				& $\sim 0$\\
$W^+W^-$			& 0 & 0			& 0 & 0	 		& 0 & 0			& 0 & 0	 			&0 & 0 				& 68 \\
\hline
\end{tabular}
\end{center}
\end{table}

For completeness, we note that the branching fraction for a technipion into a photon pair via a techniquark loop is:
\begin {equation}
{\Gamma (P \rightarrow \gamma\gamma)} = { \frac{m_{P}^3}{ 64 \pi}}   \left(\frac {\alpha_s N_{TC}{\cal A}_{\gamma \gamma}}{2 \pi F_P} \right)^2\ .
\label{eq:techni-gam}
\end {equation}
as compared with the result for the standard model Higgs boson (through a top quark loop) \cite{Gunion:1989we} 
\begin {equation}
{\Gamma (h_{SM} \rightarrow \gamma\gamma)} ={\frac{m_{h}^3}{9 \pi}}  \left(\frac{ \alpha}{3 \pi v}\right)^2  
\left[\frac{3\tau}{2}(1 + (1-\tau) f(\tau))\right]^2~,
\label{eq:higgs-gam-gam-tau}
\end {equation}

From these decay widths, we can now calculate the technipion branching ratios to all of the significant two-body final states, taking $N_{TC} = 2$ and $N_{TC} = 4$ by way of example.  In the TCSM low-scale model we set $N_D = 5\ (10)$ for $N_{TC} = 2\ (4)$ to make the technicolor coupling walk; in the Isotriplet model, we set the technifermion hypercharge to the value $y=1$.  We find that the branching ratio values are nearly independent of the size of $M_P$ within the range 110 GeV - 145 GeV and also show little variation once $M_P > 2 m_t$; to give a sense of the patterns, the branching
fractions for $M_P = 130$ GeV are shown in Table II and those for $M_P = 350$ GeV are shown in Table III.  The branching ratios for the SM Higgs at NLO are given
for comparison; these were obtained from the Handbook of LHC Higgs Cross Sections \cite{Dittmaier:2011ti}. The primary differences are the absence of a $WW$ decay for technipions and the enhancement of the two-gluon coupling (implying increased $gg \to P$ production); the di-photon and di-tau decay widths  can also vary moderately from the standard model values.

Pulling this information together, and noting that the PNGBs are narrow resonances, we may define an enhancement factor for the full production-and-decay process $yy \to {P} \to xx$ as the ratio of the products of the width of the (exclusive)
production mechanism  and the branching ratio for the decay: 
\begin{equation}
\kappa_{yy/xx}^{P} = \frac{ \Gamma({P} \to yy) \times BR({P} \to x x)}
         { \Gamma(h_{SM} \to yy) \times BR(h_{SM} \to  x x)} \equiv \kappa_{yy\ prod}\  \kappa_{xx\ decay} \ .
\label{eq:kappa}
\end{equation}
And to include both the gluon fusion and $b$-quark annihilation production channels when looking for a technipion in the specific decay channel ${P} \to xx$, we define
a combined enhancement factor 
\begin{eqnarray}
\kappa_{total/xx}^{P} 
&=&
\frac{\sigma(gg\to{P}\to xx)+\sigma(bb\to{P} \to xx)}
{{\sigma(gg\to h_{SM} \to xx)+\sigma(bb\to h_{SM} \to xx)} }\nonumber\\
&=&
\frac{\kappa_{gg/xx}^{ P}+\sigma(bb\to {P}\to xx)/\sigma(gg\to h_{SM} \to xx)}
{{1+\sigma(bb\to h_{SM} \to xx)/\sigma(gg\to h_{SM} \to xx)} }\nonumber\\
&=&
\frac{\kappa_{gg/xx}^{P} +\kappa_{bb/xx}^{P}\sigma(bb\to h_{SM} \to xx)/\sigma(gg\to h_{SM} \to xx)}
{{1+\sigma(bb\to h_{SM} \to xx)/\sigma(gg\to h_{SM} \to xx)} }\nonumber\\
&\equiv&
[\kappa_{gg/xx}^{P}+\kappa_{bb/xx}^{P} R_{bb:gg}]/
[{1+  R_{bb:gg}}]  .
\label{kappab}
\end{eqnarray}
Here 
 $R_{bb:gg}$ is the ratio of $b\bar{b}$ and $gg$ initiated Higgs boson production in the 
 Standard Model, which can be calculated using the HDECAY program \cite{Djouadi:1997yw}. In practice, as noted in \cite{Belyaev:2005ct}, the contribution from b-quark annihilation is far smaller than that from gluon fusion for colored technifermions.
 
\section{Models with colored technifermions and a topcolor mechanism}
\label{asec:topcol}

We will now show how the LHC data constrains technipions composed of colored technifermions in theories where the top-quark's mass is generated by new strong
``topcolor"  dynamics \cite{Hill:1994hp} preferentially coupled to third-generation quarks.  In such models, the ETC coupling between ordinary quarks and technifermions (or technipions) is very small, so that gluon fusion through a top-quark loop will be negligible by comparison with gluon fusion through a technifermion loop, as a source of technipion production.

\subsection{LHC Limits on Models with Light Technipions }

Here we report our results for technipions in the 110 - 145 GeV mass range where direct comparison with Higgs production is possible.  We consider final states with pairs of photons or tau leptons, since the LHC experiments have reported limits on the standard model Higgs boson in both channels.

\begin{figure}[H]
	\centering
	\subfigure[\ Original one-family model \cite{Farhi:1980xs}.]{
		\includegraphics[scale=0.65]{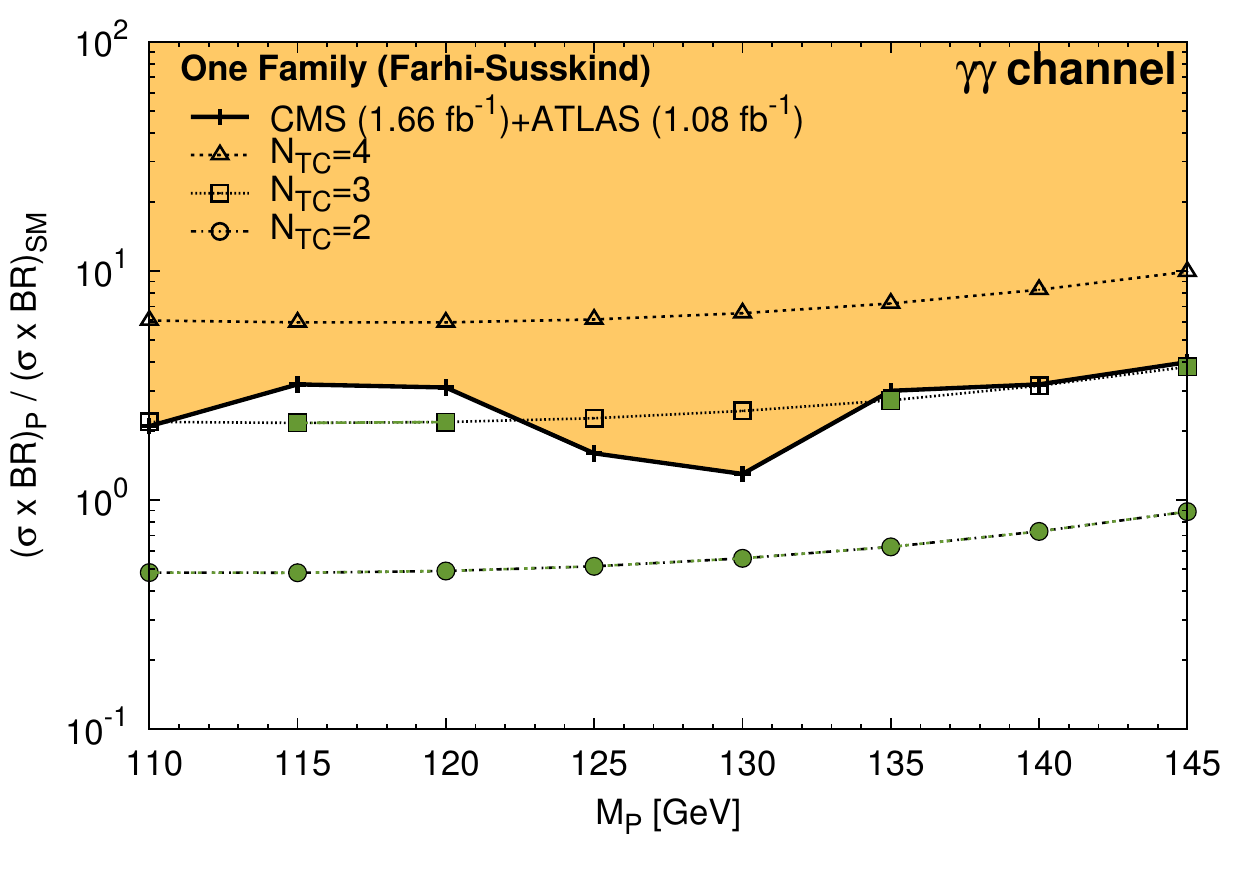}
	}
	\subfigure[\ Variant one-family model \cite{Casalbuoni:1998fs}.]{
		\includegraphics[scale=0.65]{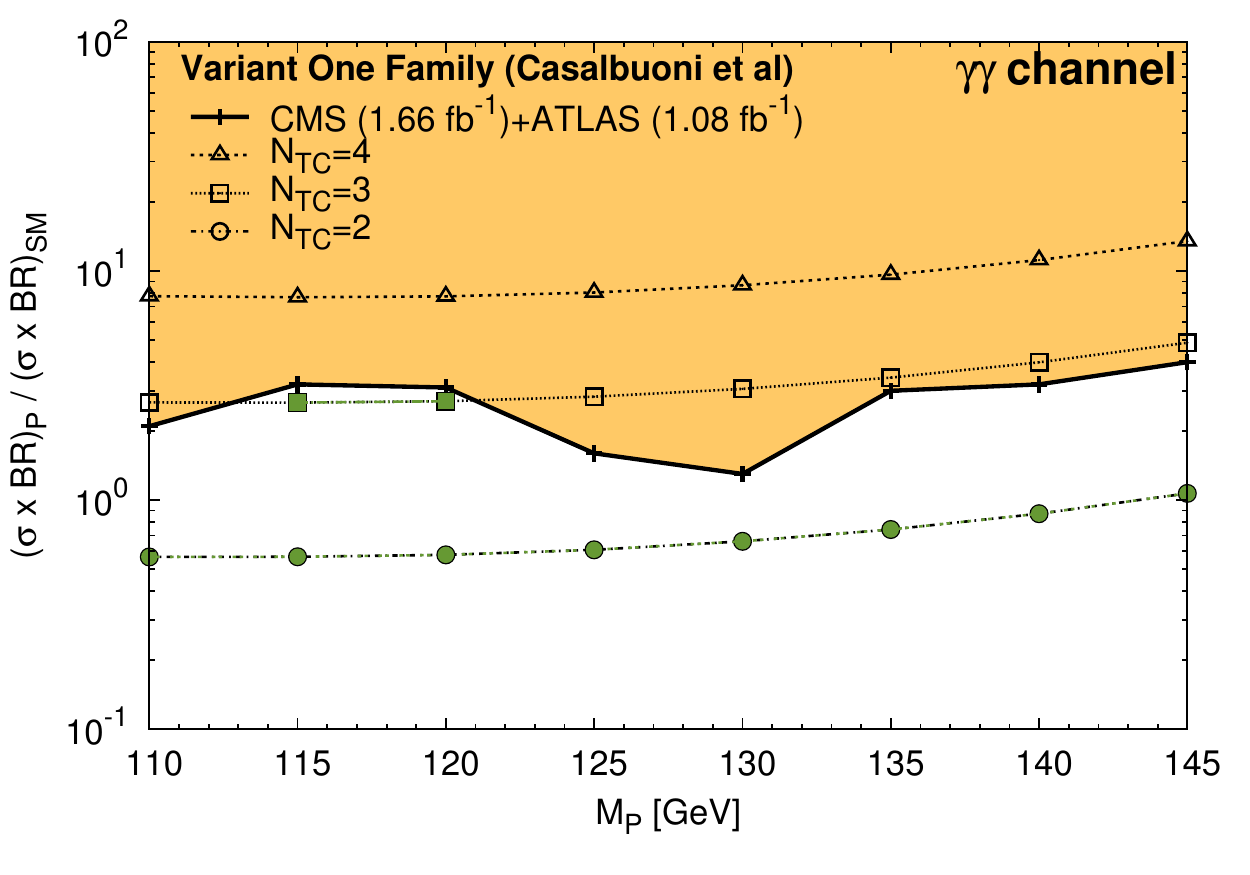}
	}
	\\
	\subfigure[\ Multiscale walking technicolor model \cite{Lane:1991qh}.]{
		\includegraphics[scale=0.65]{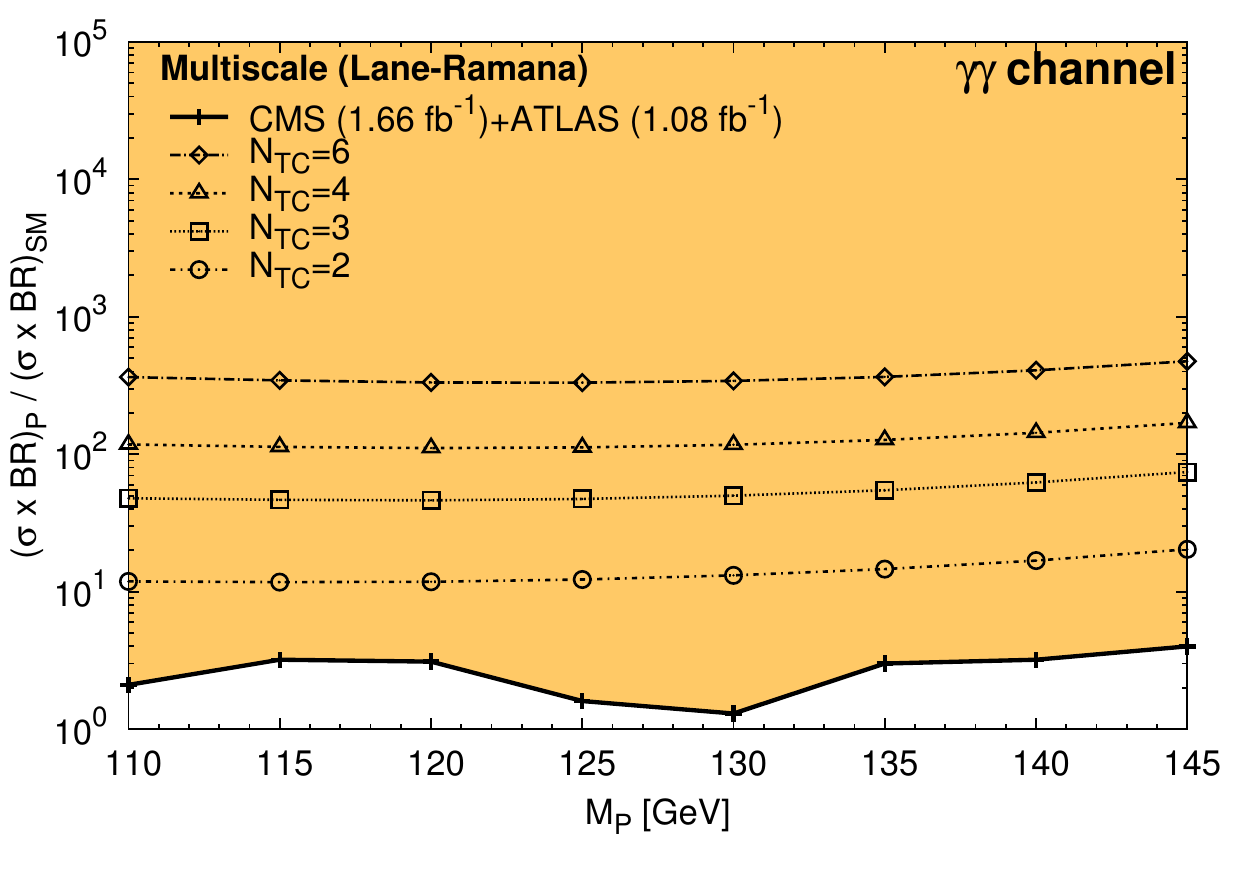}
	}
	\subfigure[\ TCSM Low-scale technciolor model (the Technicolor Straw Man model) \cite{Lane:1999uh}.]{
		\includegraphics[scale=0.65]{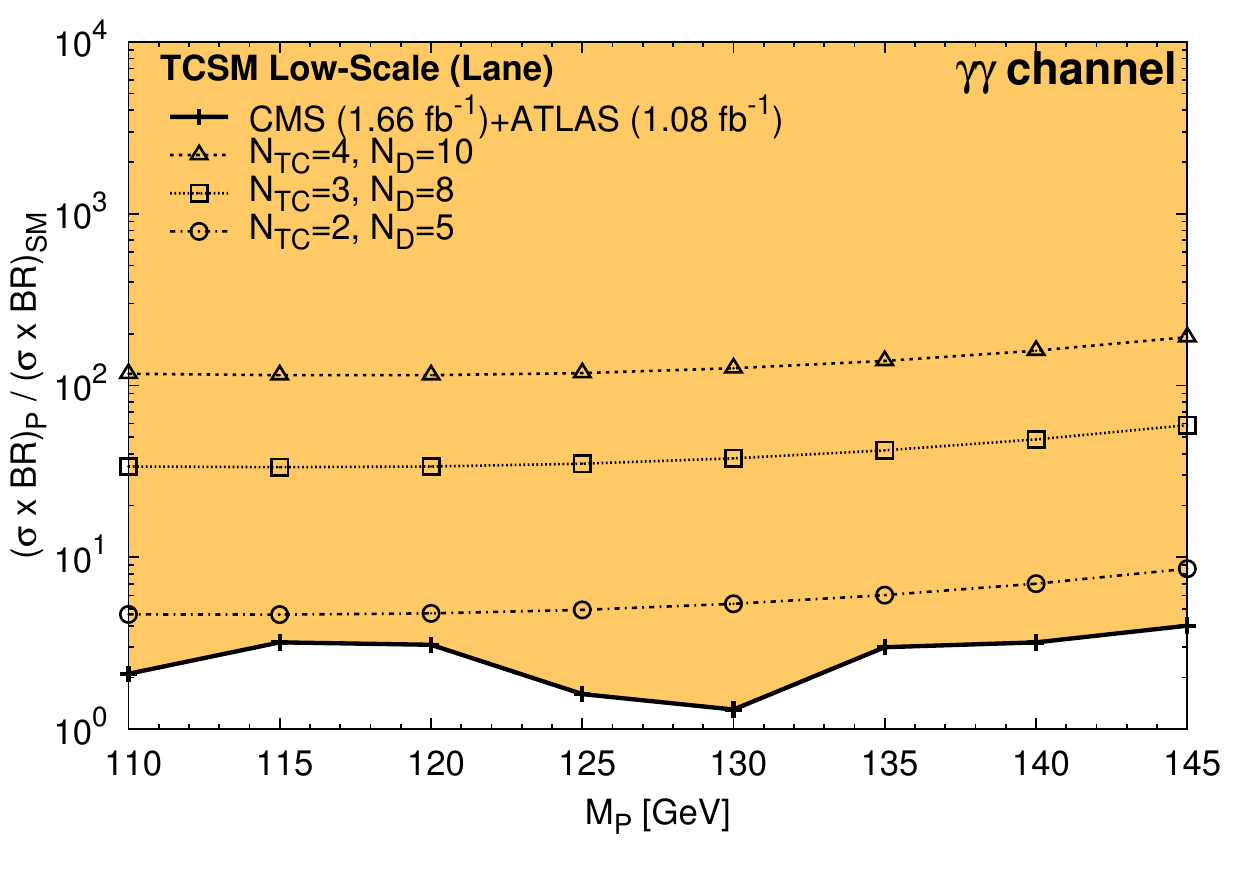}
	}
	\\	
	\subfigure[\ Isotriplet model  \cite{Manohar:1990eg}. The magnitude of the technifermion hypercharge variable $y$ has been set to 1 for illustration.]{
		\includegraphics[scale=0.65]{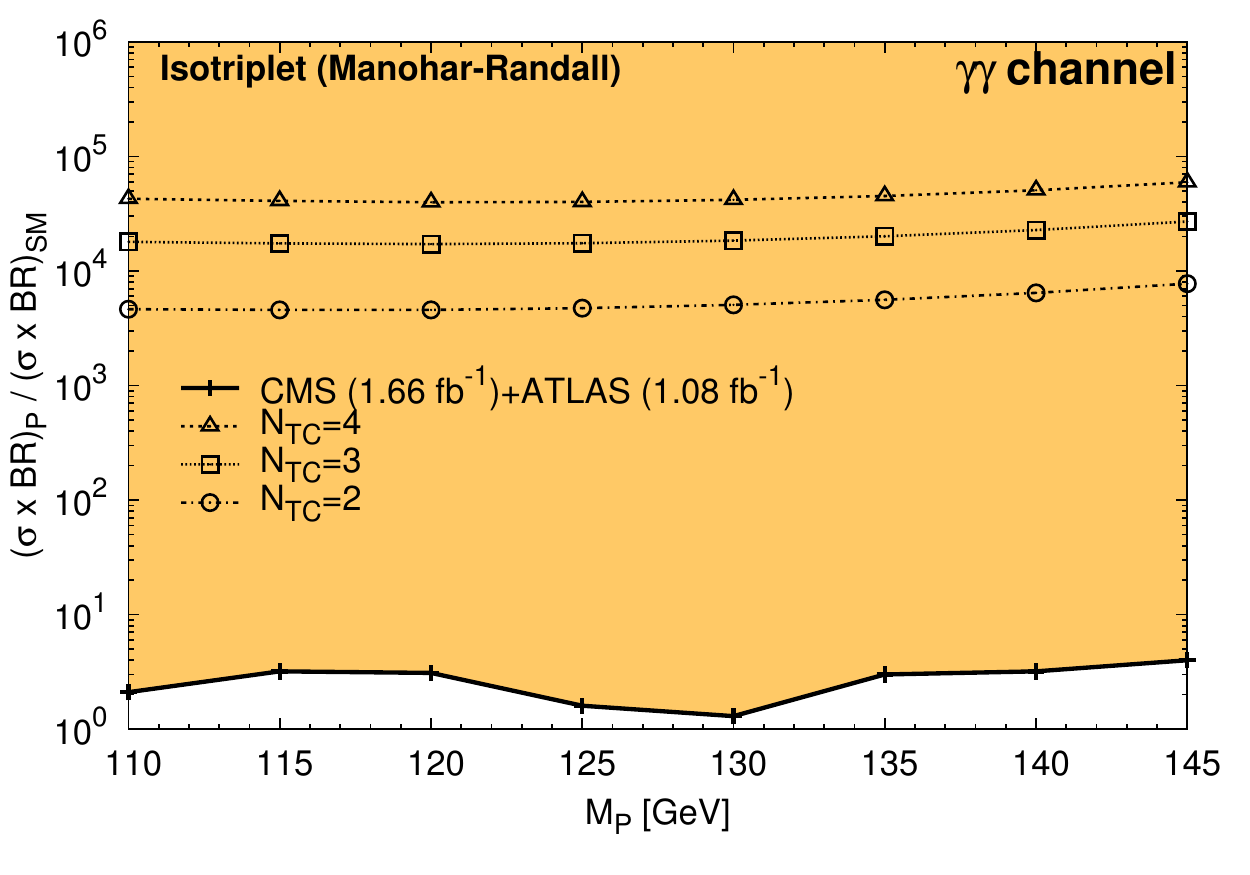}
	}

	\caption{Comparison of experimental limits and technicolor model predictions for production of a new scalar decaying to photon pairs. In each pane, the shaded region (above the solid line) is excluded by the combined $95\%$ CL upper limits on $\sigma_h B_{\gamma\gamma}$ normalized to the SM expectation as observed by CMS \cite{CMS-higgs-di-gamma} and ATLAS \cite{Collaboration:2011ww}.  Each pane also displays (as open symbols)  the theoretical prediction from one of our representative technicolor models with colored technifermions, as a function of technipion mass and for several values of $N_{TC}$.  Values of mass and $N_{TC}$ for a given model that are not excluded by the data are shown as solid (green) symbols.}
	\label{fig:HgamgamEnhancementLimit-Logscale}
	\end{figure}

First, we show the limits derived from the CMS and ATLAS searches for a standard model Higgs boson decaying to $\gamma\gamma$ in Figure \ref{fig:HgamgamEnhancementLimit-Logscale}.  The multiscale \cite{Lane:1991qh}, TCSM low-scale  \cite{Lane:1999uh}, and isotriplet \cite{Manohar:1990eg} models predict rates of technipion production and decay to diphotons that exceed the experimental limits in this mass range even for the smallest possible size of the technicolor gauge group (larger $N_{TC}$ produces a higher rate).   Note that we took the value of the technifermion hypercharge parameter $y$ in the isotriplet model to have the value $y=1$ for purposes of illustration; choosing $y \sim 1/7$ could make this model consistent with the di-photon data for $N_{TC} = 2$, but that would not affect the limits from the di-tau channel discussed below.   For the original \cite{Farhi:1980xs} and variant  \cite{Casalbuoni:1998fs} one-family models, the data still allow $N_{TC} = 2$ over the whole mass range, and $N_{TC} = 3$ is possible for 115 GeV $< M_P <$ 120 GeV; even 135 $< M_P <$ 145 GeV is marginally consistent with the data for $N_{TC}=3$ in the original one-family model.

	\begin{figure}[H]
	\centering
	\subfigure[\ Original one-family model \cite{Farhi:1980xs}.]{
		\includegraphics[scale=0.65]{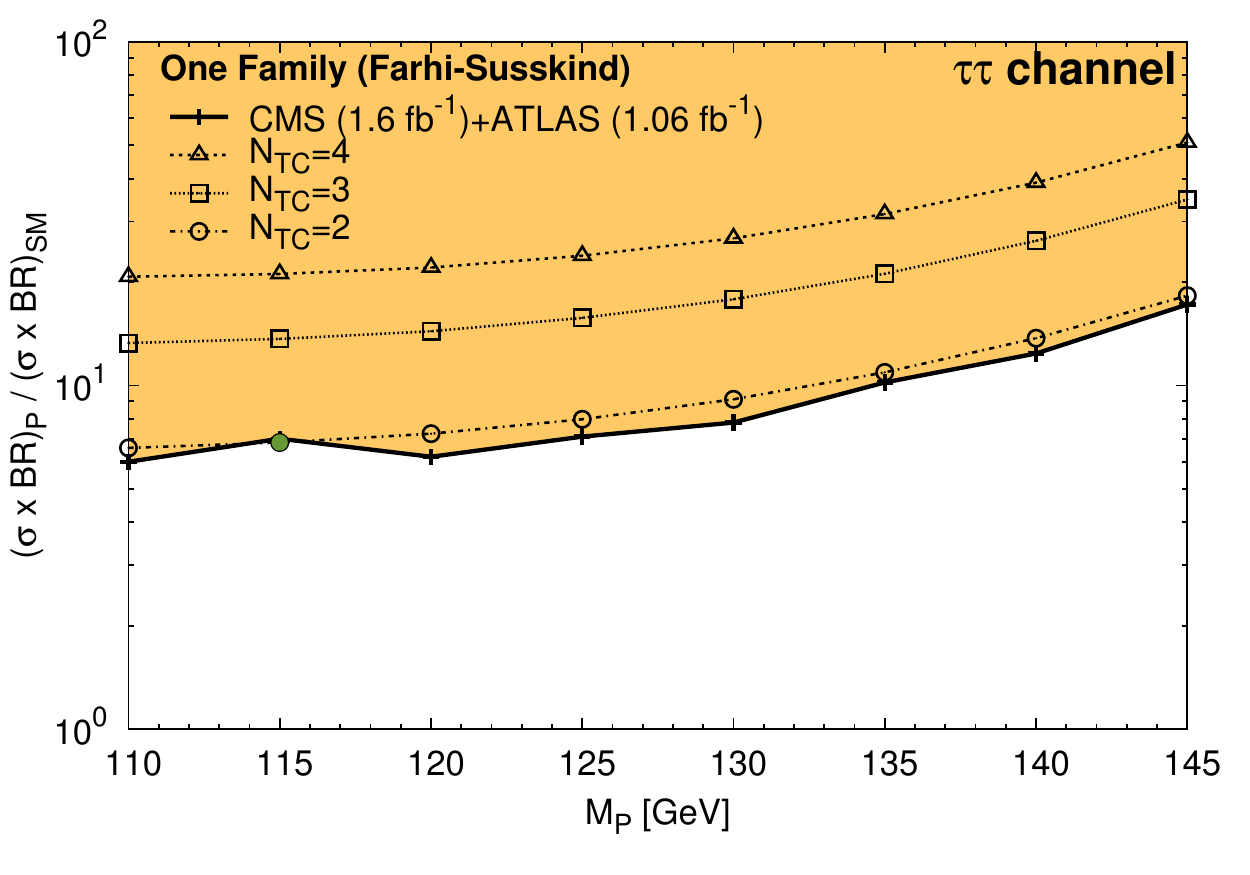}
	}
	\subfigure[\ Variant one-family model \cite{Casalbuoni:1998fs}.]{
		\includegraphics[scale=0.65]{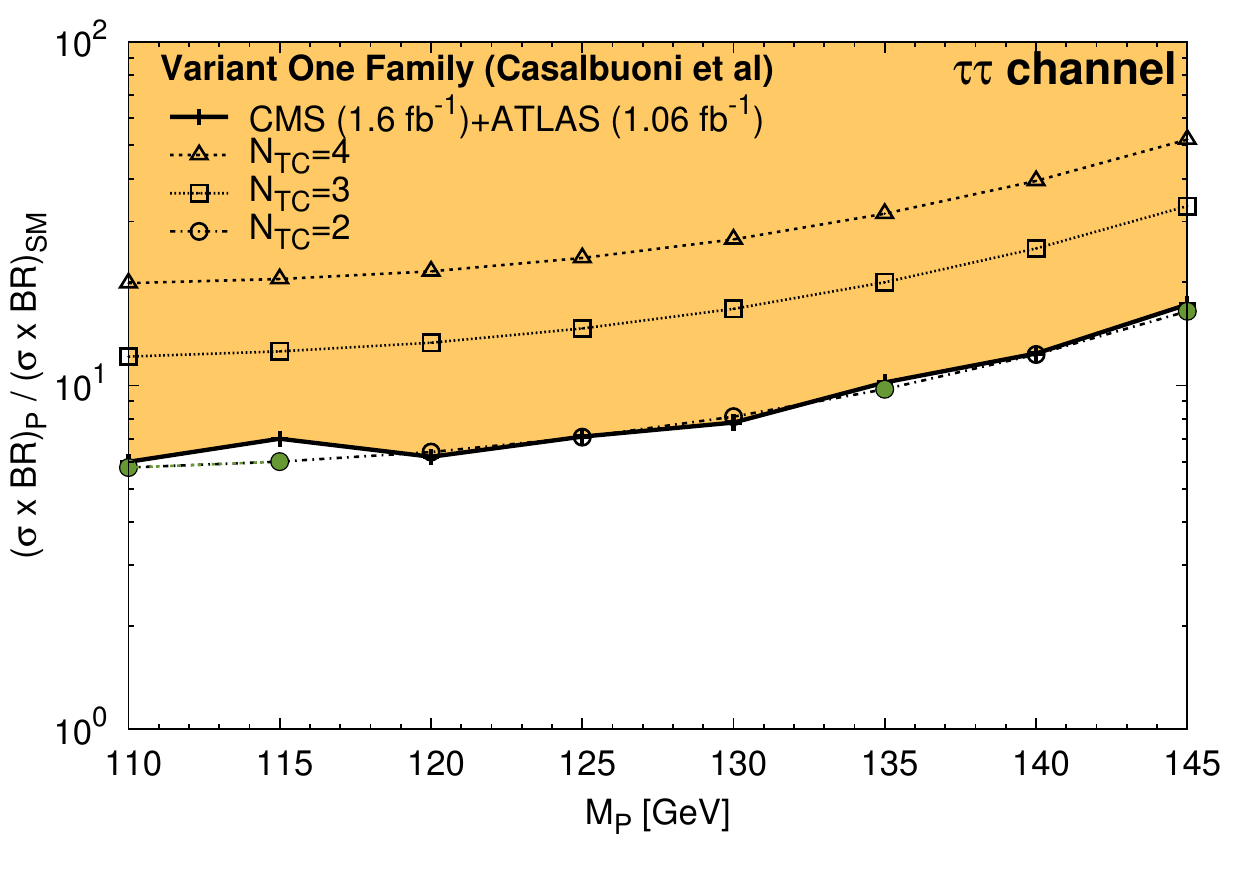}
	}
	\\
	\subfigure[\ Multiscale walking technicolor model \cite{Lane:1991qh}.]{
		\includegraphics[scale=0.65]{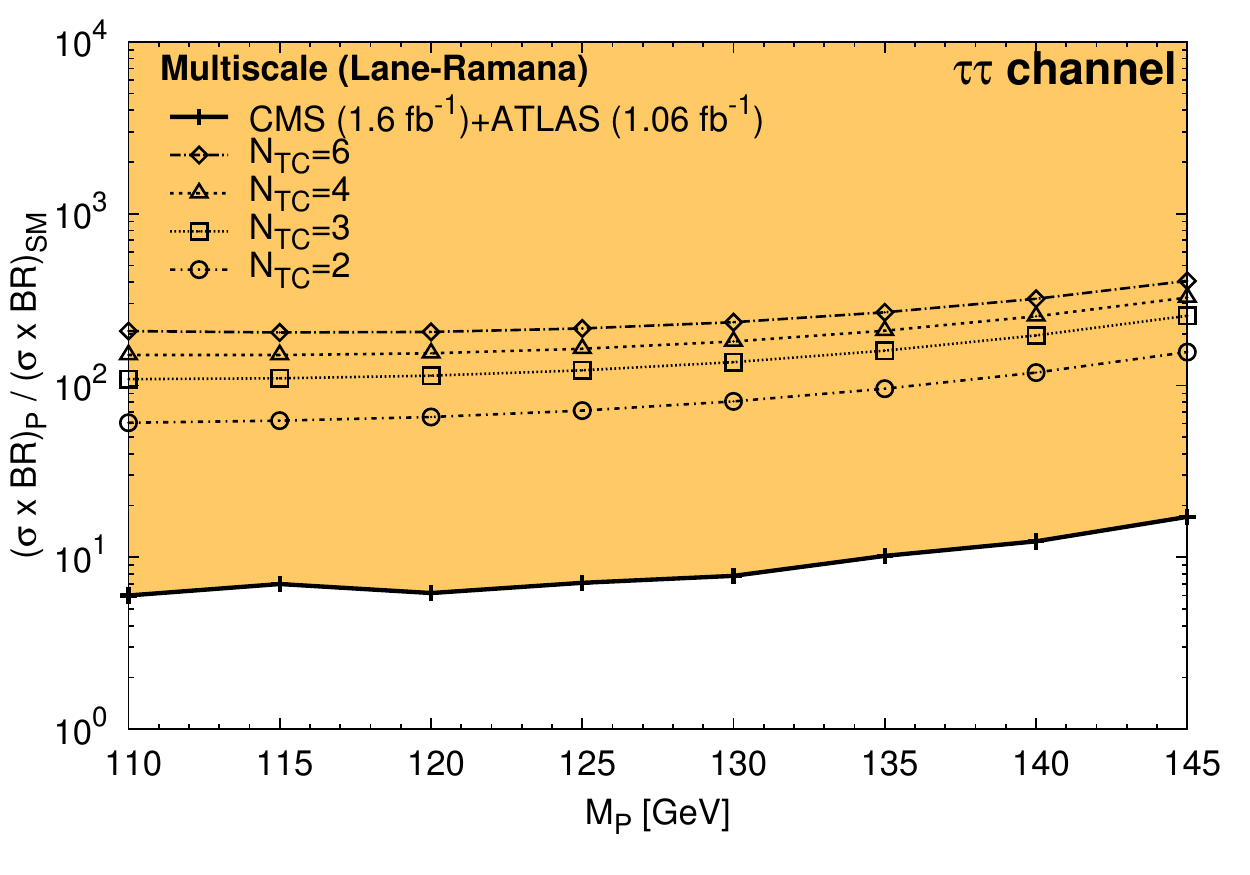}
	}
	\subfigure[\ TCSM Low-scale technciolor model (the Technicolor Straw Man model) \cite{Lane:1999uh}.]{
		\includegraphics[scale=0.65]{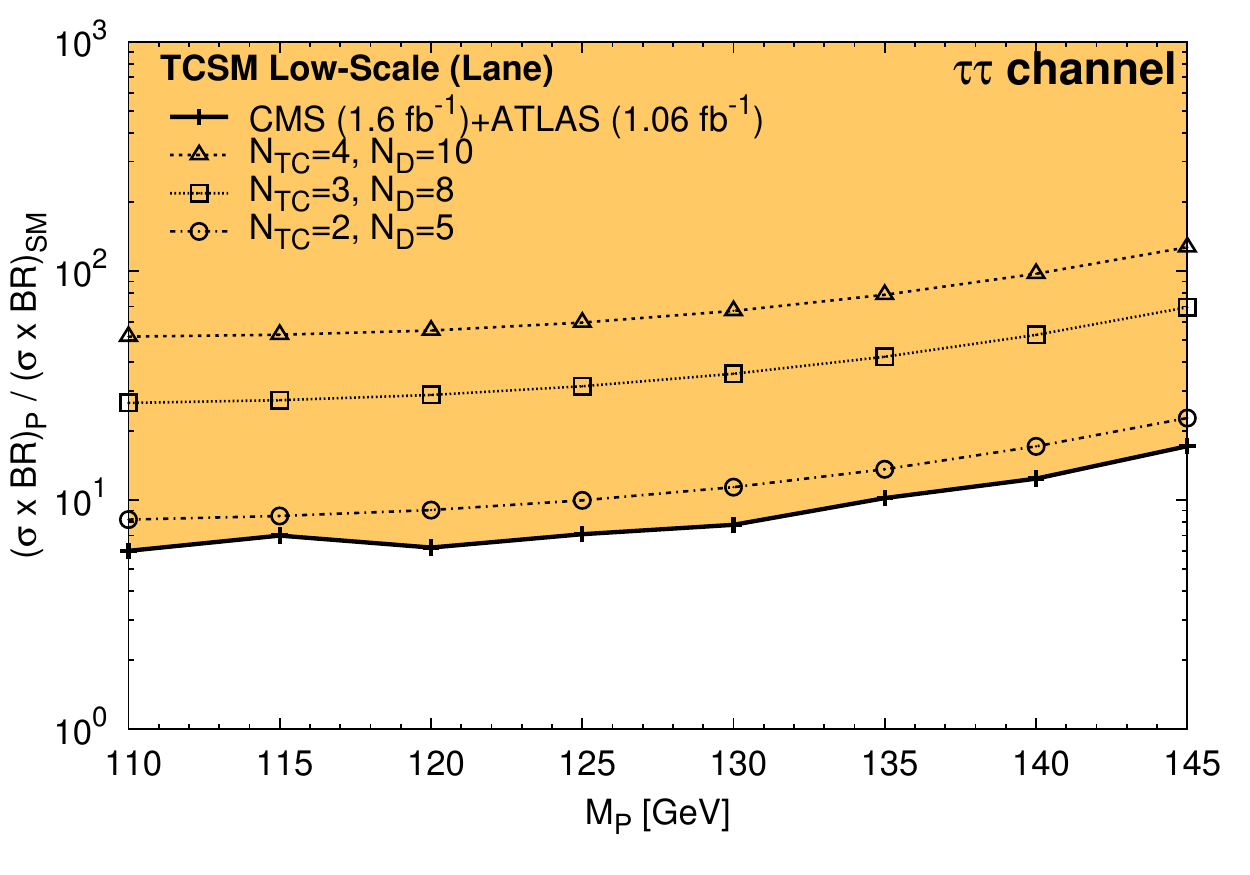}
	}
	\\	
	\subfigure[\ Isotriplet model \cite{Manohar:1990eg}.The magnitude of the technifermion hypercharge variable $y$ has been set to 1 for illustration]{
		\includegraphics[scale=0.65]{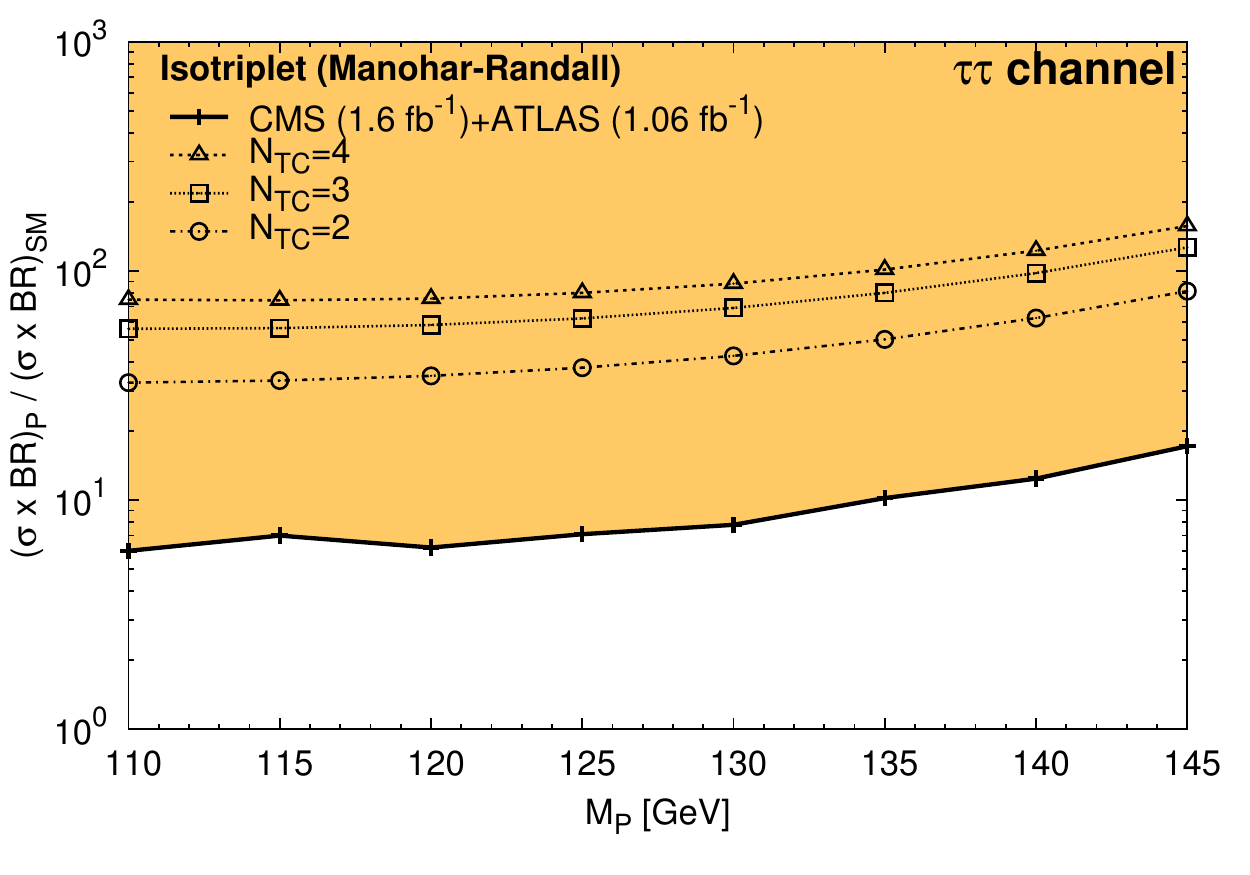}
	}

	\caption{Comparison of experimental limits and technicolor model predictions for production of a new scalar decaying to tau lepton pairs. In each pane, the shaded region (above the solid line) is excluded by the combined $95\%$ CL upper limits on $\sigma_h B_{\tau^+\tau^-}$ normalized to the SM expectation as observed by CMS \cite{CMS-higgs-di-tau} and ATLAS \cite{ATLAS-higgs-di-tau-low}.  Each pane also displays (as open symbols)  the theoretical prediction from one of our representative technicolor models with colored technifermions, as a function of technipion mass and for several values of $N_{TC}$.  Values of $M_P$ and $N_{TC}$ for a given model that are not excluded by the data are shown as solid (green) symbols; the only such point is at $N_{TC}= 2$ and $M_P = 115$ GeV for the variant one-family model.}
	\label{fig:HtautauEnhancementLimit-Logscale}
	\end{figure}

The limits from the the CMS and ATLAS searches for a standard model Higgs boson decaying to $\tau^+\tau^-$ in the same mass range are even more stringent, as shown in Figure  \ref{fig:HtautauEnhancementLimit-Logscale}.  The data again exclude the multiscale \cite{Lane:1991qh}, TCSM low-scale  \cite{Lane:1999uh}, and isotriplet \cite{Manohar:1990eg} models across the full mass range and for any size of the technicolor gauge group.  The original \cite{Farhi:1980xs} is likewise excluded; only $M_P = 115$ GeV for $N_{TC} = 2$ is even marginally consistent with data.  The variant  \cite{Casalbuoni:1998fs} one-family model is marginally consistent with data for $N_{TC} = 2$ but excluded for all higher values of $N_{TC}$.  Forthcoming LHC data on $\tau\tau$ final states should provide further insight on these two models for $N_{TC} = 2$.

\subsection{LHC Limits on Heavier Technipions Decaying to Tau-Lepton Pairs}

\begin{figure}[H]
	\centering
	\subfigure[\ Original one-family model \cite{Farhi:1980xs}.]{
		\includegraphics[scale=0.65]{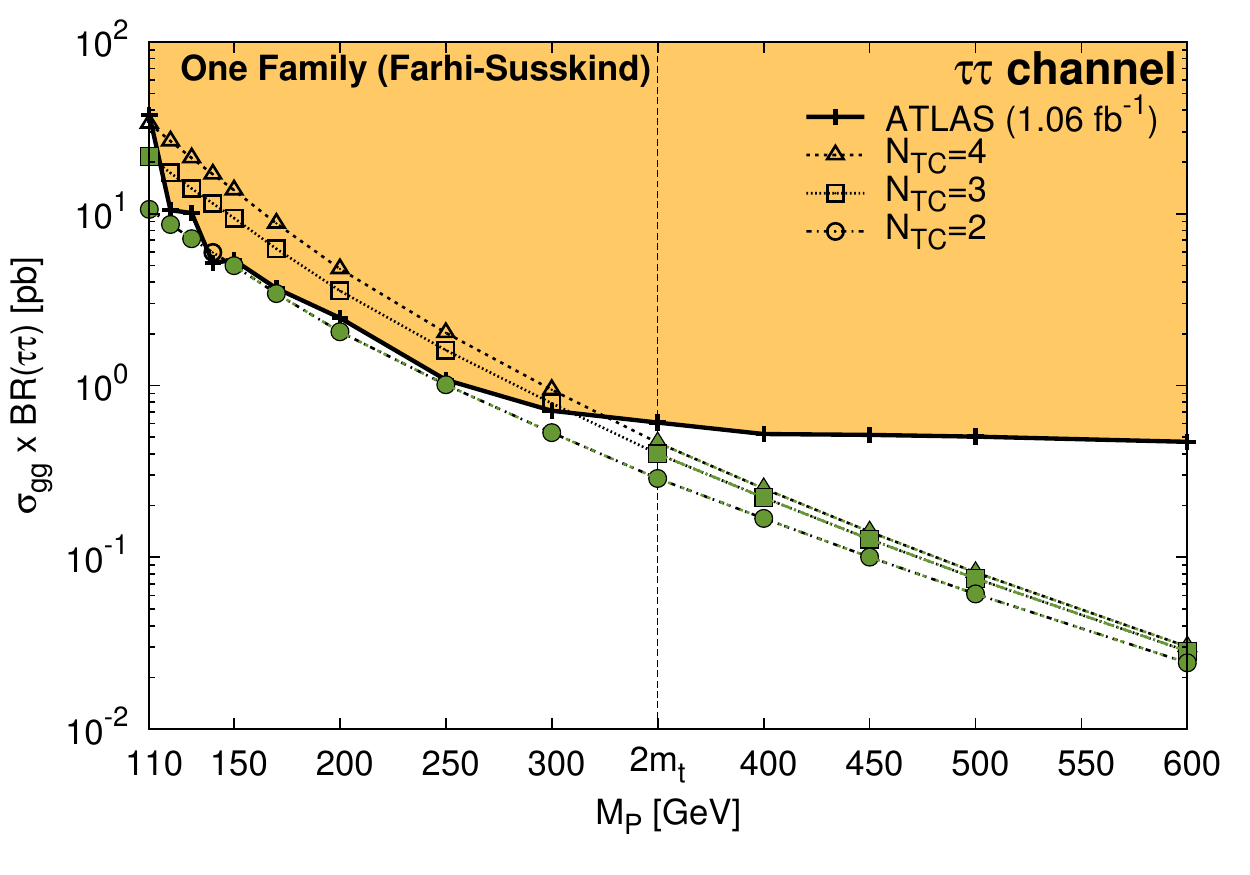}
	}
	\subfigure[\ Variant one-family model \cite{Casalbuoni:1998fs}.]{
		\includegraphics[scale=0.65]{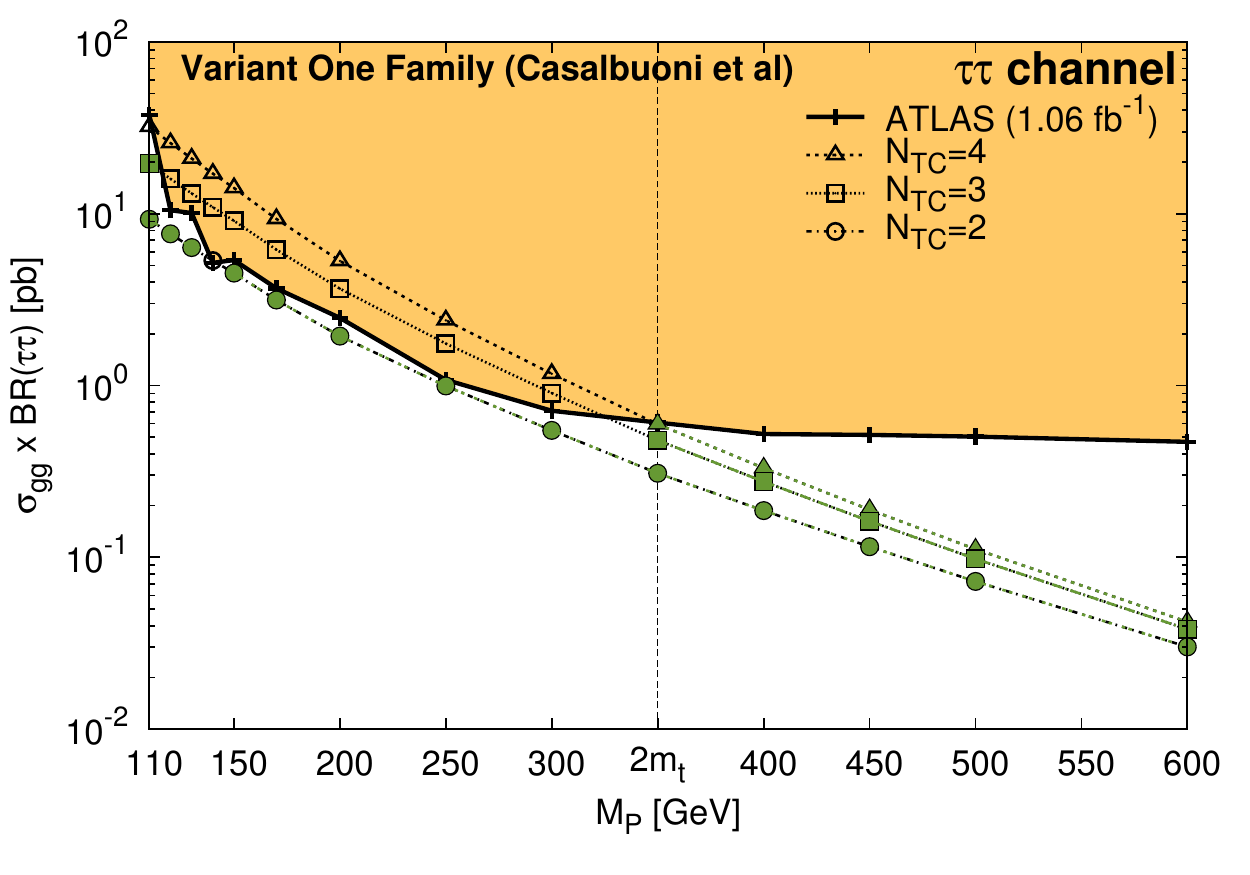}
	}
	\\
	\subfigure[\ Multiscale walking technicolor model \cite{Lane:1991qh}.]{
		\includegraphics[scale=0.65]{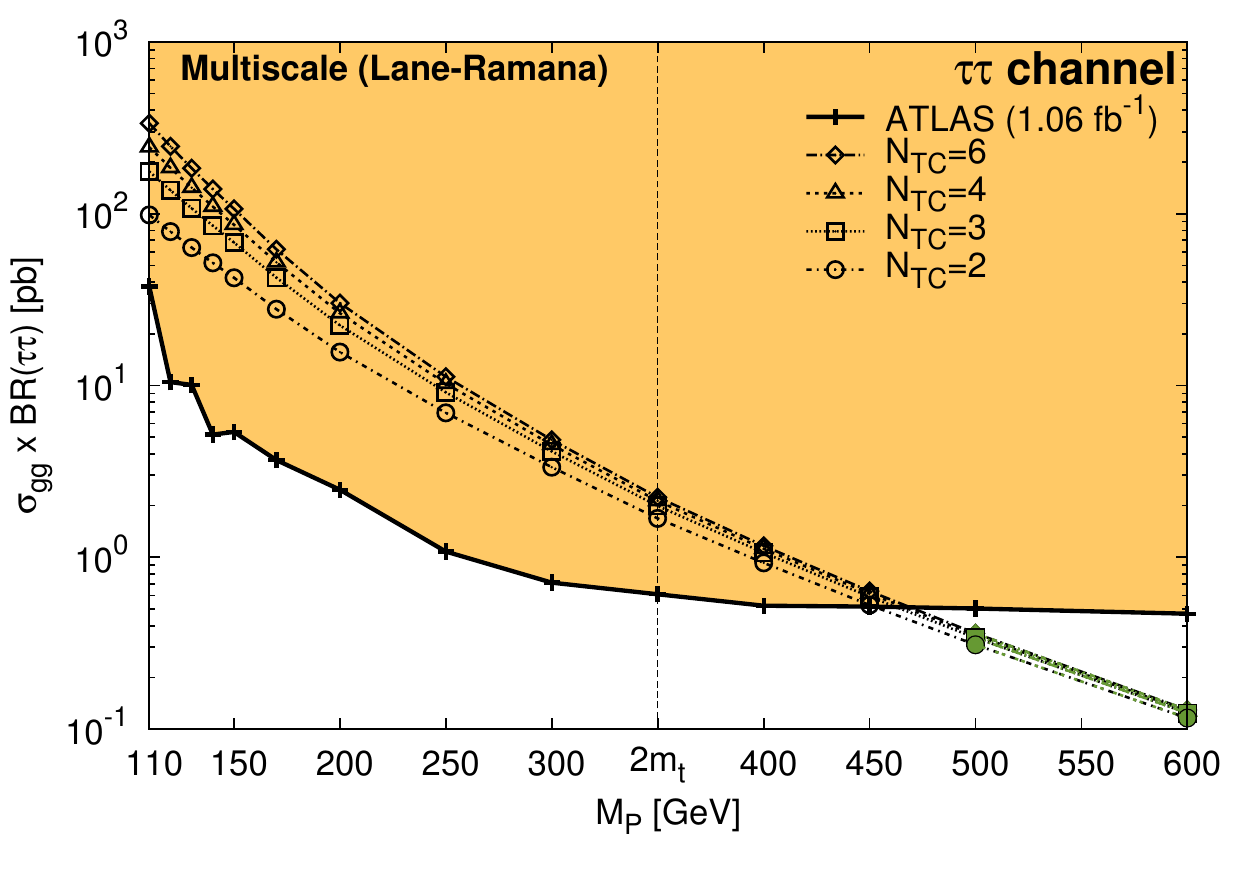}
	}
	\subfigure[\ TCSM Low-scale technciolor model (the Technicolor Straw Man model) \cite{Lane:1999uh}.]{
		\includegraphics[scale=0.65]{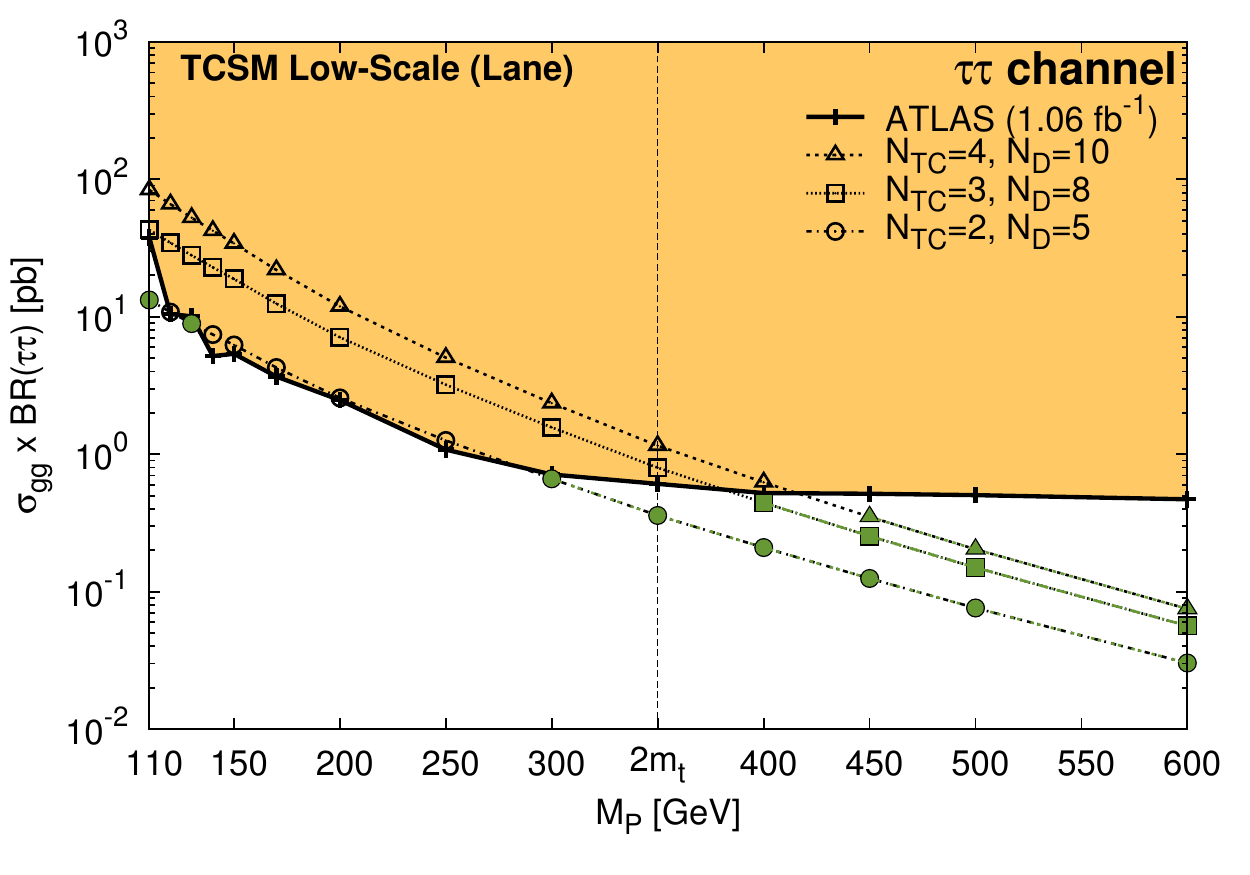}
	}
		\\	
	\subfigure[\ Isotriplet model \cite{Manohar:1990eg}.The magnitude of the technifermion hypercharge variable $y$ has been set to 1 for illustration]{
		\includegraphics[scale=0.65]{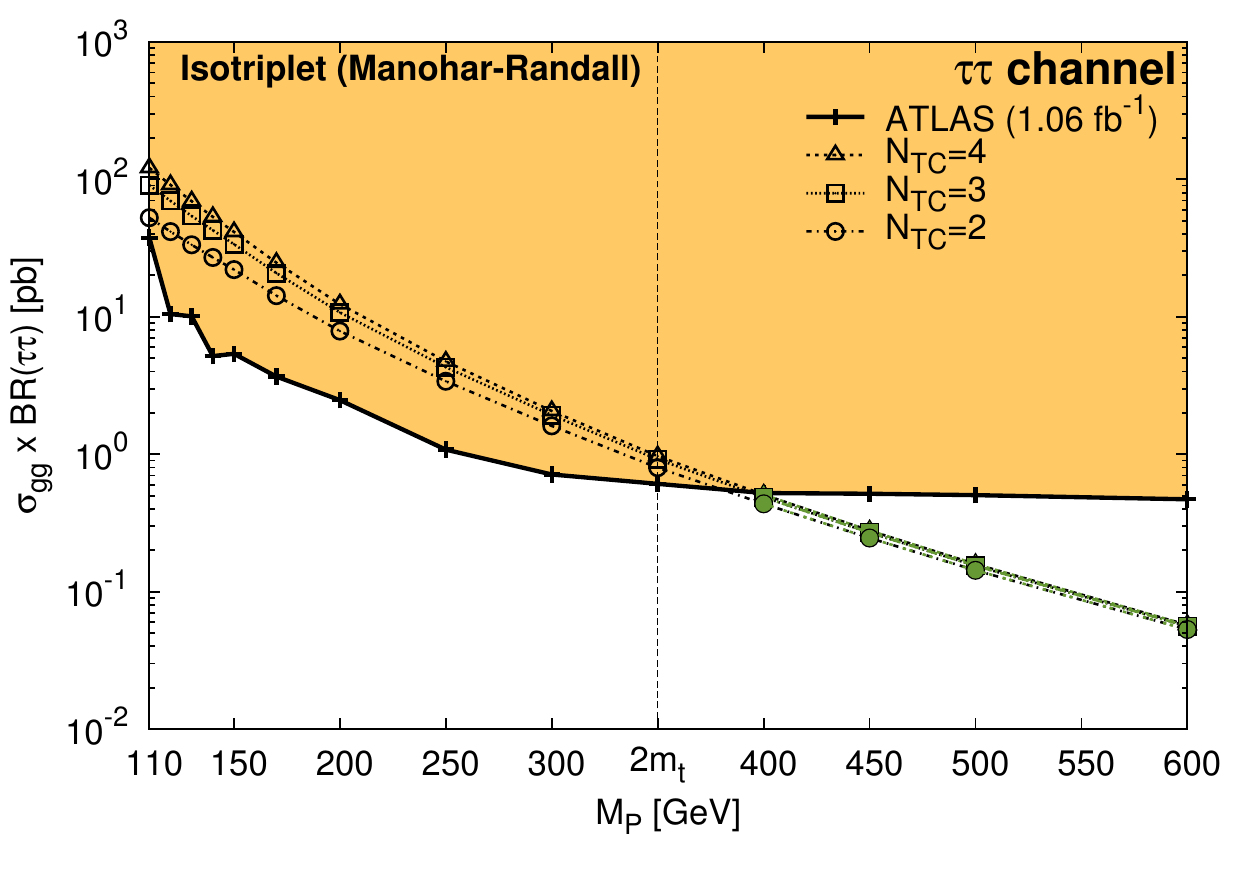}
	}

	\caption{Comparison of experimental limits and technicolor model predictions for production of a new scalar decaying to tau lepton pairs for scalar masses in the mass range 110 - 600 GeV . In each pane, the shaded region (above the solid line) is excluded by the  $95\%$ CL upper limits on $\sigma_h B_{\tau^+\tau^-}$ from ATLAS  \cite{ATLAS-higgs-di-tau-high}.  Each pane also displays (as open symbols)  the theoretical prediction from one of our representative technicolor models with colored technifermions, as a function of technipion mass and for several values of $N_{TC}$.  Values of $M_P$ and $N_{TC}$ for a given model that are not excluded by this data are shown as solid (green) symbols; nearly all such values at low technipion masses are excluded by the data shown in Figure 2. As discussed in the text, limits to the right of the vertical bar at $M_P = 2 m_t$ apply only when a topcolor sector, rather than extended technicolor, generates most of the top quark's mass.}
	\label{fig:XSBRggHtautauATLASTau-log}
	\end{figure}

We now consider technipions that are too heavy to be directly compared with a Higgs in the LHC data, but which can be directly constrained by looking at data from final states with tau-lepton pairs.  ATLAS has obtained \cite{ATLAS-higgs-di-tau-high}  limits on the product of the production cross section with the branching ratio to tau pairs at $95\%$ confidence level for a generic scalar boson in the mass range $100-600$ GeV.   We use this limit to constrain technicolor models as follows.   The production cross section $\sigma(gg\rightarrow P)$ for technicolor models can be estimated by scaling from the standard model\footnote{The standard model production cross section $ \sigma(gg\rightarrow h_{SM})$ at several values of the Higgs mass can be obtained from the Handbook \cite{Dittmaier:2011ti}. } using the production enhancement factor calculated for each technicolor model \cite{Belyaev:2005ct}.  And the branching fraction of the technipions into tau pairs is shown in Table II, above.  Therefore,
	\begin{eqnarray}
	\sigma(gg\rightarrow P){BR(P\rightarrow\tau\tau)} &=& \kappa_{gg \  prod} \sigma(gg\rightarrow h_{SM}) BR(P \rightarrow \tau\tau)\, .
	\end{eqnarray}
Our comparison of the experimental limits with the model predictions is shown in figure \ref{fig:XSBRggHtautauATLASTau-log}.  

In the region of the figures to the left of the vertical bar, we see that the data excludes technipions in the mass range from 145 GeV up to nearly $2m_t$ in all models for $N_{TC} \geq 3$.  For the multiscale and isotriplet models, $N_{TC} = 2$ is excluded as well in this mass range; for the TCSM low-scale model, $N_{TC} = 2$ is excluded up to nearly 300 GeV (the few points that are allowed at low mass on this plot are excluded by the data shown in Figures 2 and 3); while for the original and variant one-family models, $N_{TC} = 2$ can be consistent with data at these higher masses.   Again, further LHC data on di-tau final states will be valuable for discerning whether the models with only two technicolors remain viable.  At present, technicolor models with colored technifermions are strongly constrained even if their lightest technipion is just below the threshold at which it can decay to top-quark pairs.

Moreover, as the region of the figures to the right of the vertical bar demonstrates, the data also impacts technipions in the mass range above $2 m_t$ in some cases:  $M_P \leq 450$ GeV (375 GeV) is excluded for any size technicolor group in the multiscale (isotriplet) model and $M_P \leq 375$ GeV is excluded for $N_{TC} \geq 3$ in the TCSM low-scale model.  Note that these limits apply only in cases where the technipion has a very small branching fraction into top quarks, and the branching fraction to di-taus just varies smoothly with the increasing mass of the technipion.   As we shall discuss shortly, such limits on technipions heavier than $2 m_t$ would not hold in models where the extended technicolor dynamically generates the bulk of the top quark mass and the technipion has an appreciable top-quark branching fraction.

\section{Models with colored technifermions and a top mass generated by ETC}
\label{asec:topetc}

\begin{figure}[H]
	\centering
	\subfigure[\ Original one-family model \cite{Farhi:1980xs}.]{
		\includegraphics[scale=0.65]{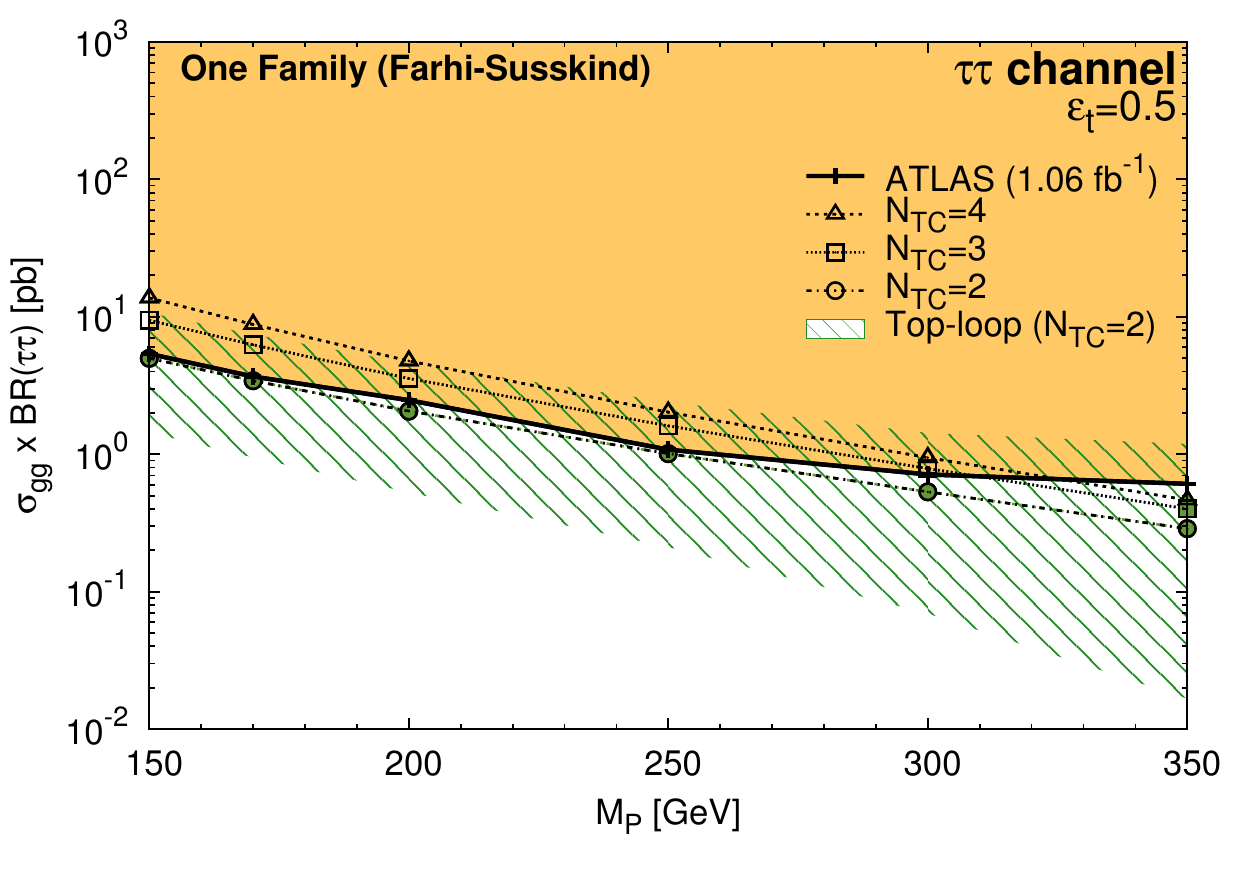}
	}
	\subfigure[\ Variant one-family model \cite{Casalbuoni:1998fs}.]{
		\includegraphics[scale=0.65]{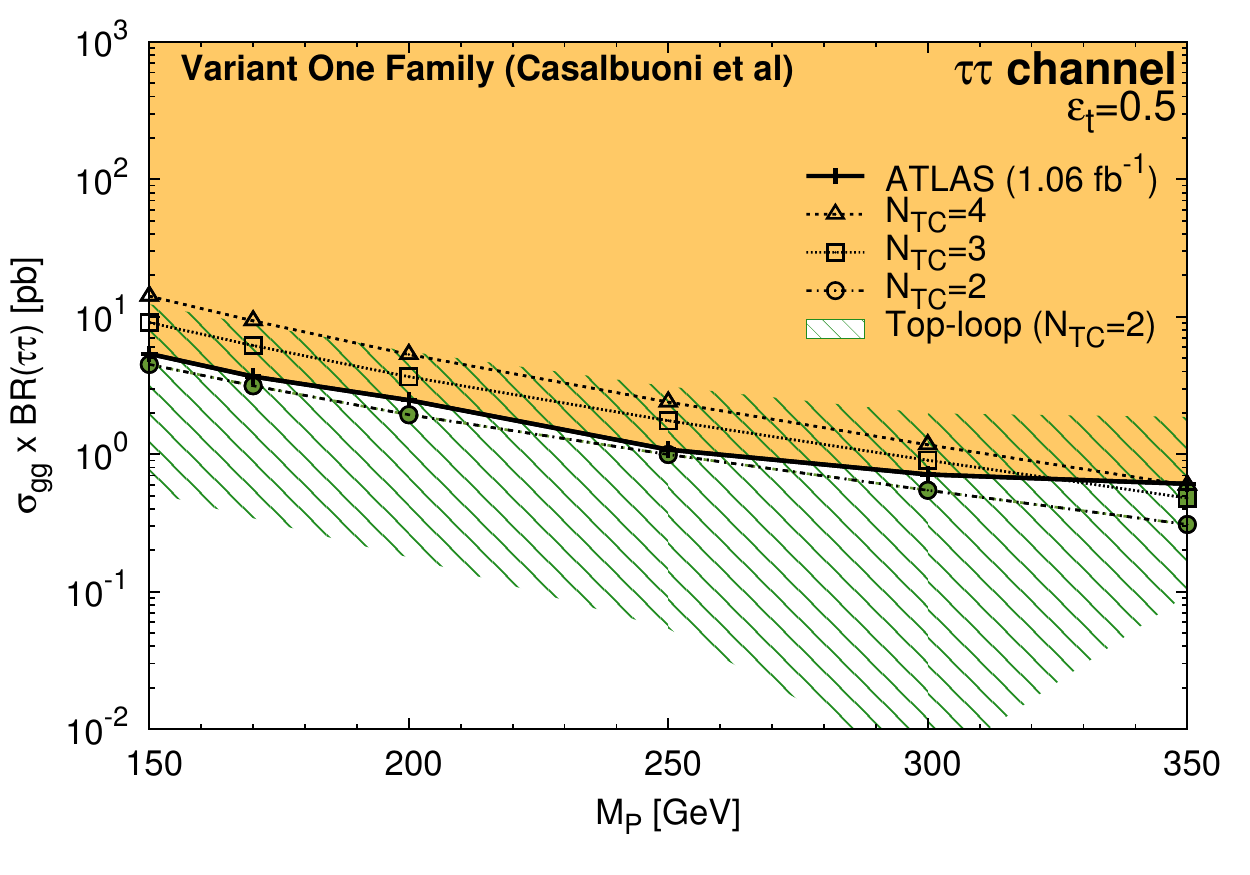}
	}
	\\
	\subfigure[\ Multiscale walking technicolor model \cite{Lane:1991qh}.]{
		\includegraphics[scale=0.65]{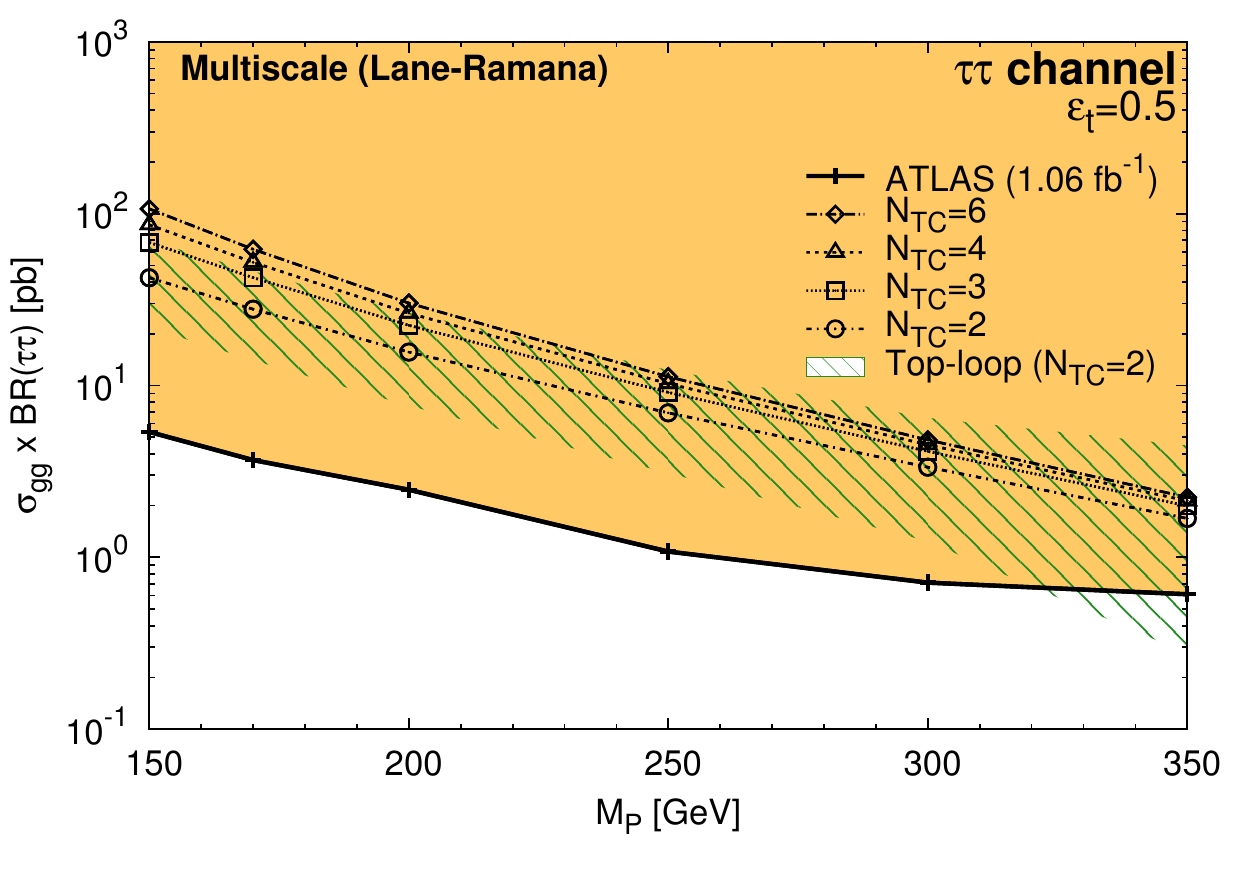}
	}
	\subfigure[\ TCSM Low-scale technciolor model (the Technicolor Straw Man model) \cite{Lane:1999uh}.]{
		\includegraphics[scale=0.65]{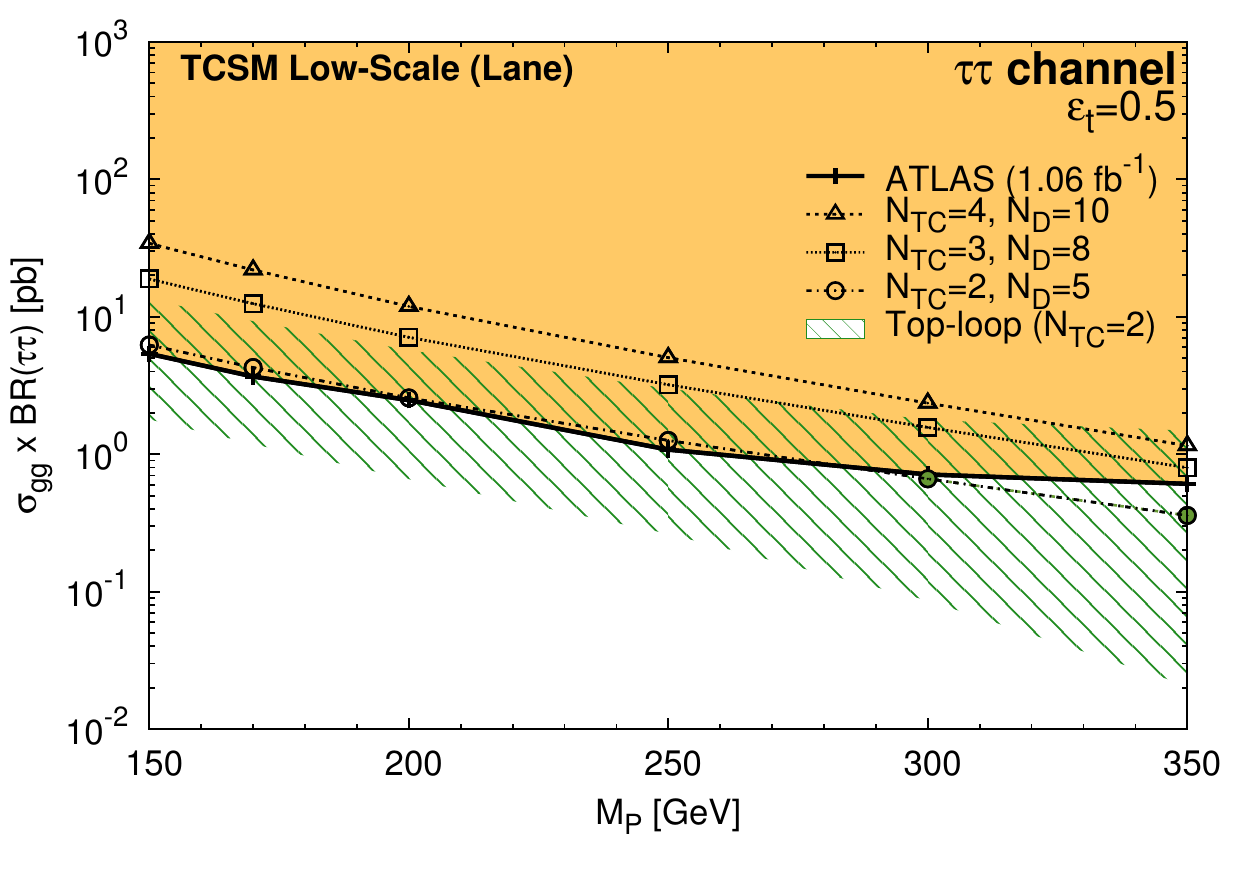}
	}
	\\	
	\subfigure[\ Isotriplet model \cite{Manohar:1990eg}.]{
		\includegraphics[scale=0.65]{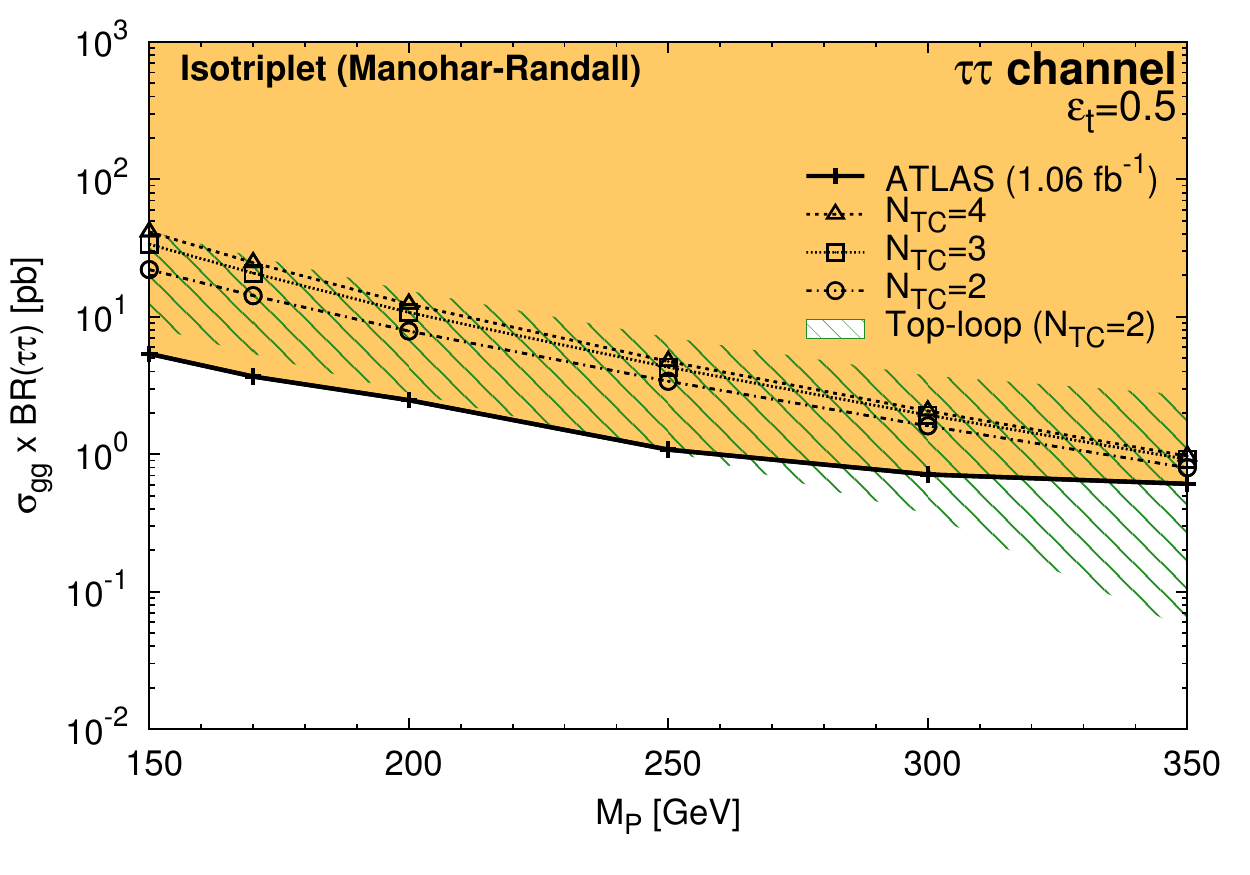}
	}

	\caption{Comparison of data and theory for production of a new scalar of mass 150 - 350 GeV that decays to tau lepton pairs; here, technipion production through techniquark loops is potentially modified by including production via top quark loops assuming extended technicolor generates most of the top quark's mass.  In each pane, the shaded region (above the solid line) is excluded by the  $95\%$ CL upper limits on $\sigma_h B_{\tau^+\tau^-}$ from ATLAS  \cite{ATLAS-higgs-di-tau-high}.  As in Figure 4, each pane displays the theoretical prediction (including techniquark loops only) from one technicolor model with colored technifermions, as a function of technipion mass and for several values of $N_{TC}$.  Values of $M_P$ and $N_{TC}$ for a given model that are not excluded by this data are shown as solid (green) symbols.  The hatched region indicates (for $N_{TC} = 2$)  how including the contributions of top-quark loops could impact the model prediction, assuming $\epsilon_t = 0.5$.  If the top and techniquark loop contributions interfere constructively, the model prediction moves to the top of the hatched region; if they interfere destructively, the model prediction moves to the bottom of the hatched region.}
	\label{fig:XSBRggHtautauATLASTau-log-lambda-0-5}
	\end{figure}

We will now illustrate how the above constraints are modified in theories where the top-quark's mass includes a substantial contribution from extended technicolor.  In such models, the ETC coupling between the top quark and technipion can be relatively large, which has several consequences.

First, it means that for technipions heavy enough to decay to top-quark pairs that channel will dominate, so that the branching fractions to $\tau^+\tau^-$ and $\gamma\gamma$ become negligible.  So these models can be constrained by the LHC data discussed in this paper only for $M_P < 2 m_t$.  Second, it implies that charged technipions $P^+$ that are lighter than the top quark can open a new top-quark decay path:  $t \to P^+ b$.   Existing bounds on this decay rate preclude charged technipions lighter than about 160 GeV; for simplicity, we will take this to be an effective lower bound on the mass of our neutral technipions in our discussion here -- though, in principle, it is possible for the neutral technipion to be lighter than its charged counterpart.   Based on these considerations, we will be considering possible LHC bounds on technipions with substantial coupling to top quarks and lying in the mass range 160 GeV $ < m_P < 2 m_t$; at present only data on di-tau final states exists for this mass range.  

Within this mass range, the presence of a large top-technipion coupling allows gluon fusion through a top-quark loop (as in Figure 1(c)) to become a significant source of technipion production.  Extrapolating from the expressions for decay of a pseudoscalar boson in \cite{Gunion:1989we}, one finds that the decay of technipion $P$ to gluons through a top-quark loop has the rate:
\begin{equation}
{\Gamma^{top} (P \rightarrow gg)} = { \frac{m_{P}^3}{ 8 \pi}}   \left(\frac {\alpha_s \epsilon_t}{2 \pi F_P} \right)^2 [\tau f(\tau)]^2
\end {equation}
where $\epsilon_t$ is the ETC-mediated top-quark coupling to technipions, $\tau$ and $f(\tau)$ are as defined in Eqs. (\ref{eq:higgs-tau}) and (\ref{eq:fftau}), and the expression $[\tau f(\tau)]\rightarrow 1$ in the limit of large top-quark mass.  Comparing this with Eqn. (\ref{eq:techni-glu}), we see that the ratio
\begin{equation}
\frac{\Gamma^{top} (P \rightarrow gg)} {\Gamma (P \rightarrow gg)} = \left( \frac{\epsilon_t [ \tau f(\tau)] }{N_{TC} {\cal A}_{gg}}\right)^2 \equiv \left( R^{loops} \right)^2
\end{equation}
can be substantial if $\epsilon_t \approx 1$ and $N_{TC}$ is small. 

The relative sign of the techniquark loop and top-quark loop contributions depends on the structure of the ETC sector of the theory.  In models where this sign is positive, the top-quark and techniquark amplitudes will add constructively and the limits derived in the previous section will be strengthened.  However, in models where the relative sign is negative, the diagrams in Figure 1(a) and 1(c) will interefere destructively,  reducing the rate of technipion production calculated in the previous section by a factor of
\begin{equation}
\left(1 - R^{loops}\right)^2
\end{equation}
That has the potential to weaken the bounds from the LHC data.

Moreover, in a technicolor model where both $N_{TC}$ and  ${\cal A}_{gg}$ are relatively small, for light technipion masses where $\tau f(\tau) \approx 1$, the ratio $R^{loops}$ can be greater than one, meaning that the top-quark loop can contribute more to technipion production than the techniquark loop.  For heavier technipion masses, the relative importance of the top-quark loop declines, and the two contributions interfere strongly, so that the production rate declines and the limits from LHC data become much weaker.  For still heavier technipion masses, the techniquark loop begins to dominate again and the interference loses its impact on the strength of the bounds.

This behavior is visible in Figure \ref{fig:XSBRggHtautauATLASTau-log-lambda-0-5}, which shows how the limits on the $N_{TC} = 2$ version of each model would be affected by the presence of top-quark loops with $\epsilon_t = 0.5$.  The data, shaded region, and model prediction curve are as in Figure \ref{fig:XSBRggHtautauATLASTau-log}, for the mass range 160 GeV $< M_P < 2 m_t$.  Also shown here is a hatched region that illustrates how the model curve would move upwards (downwards) in the presence of constructive (destructive) interference between the top and technifermion loops.  The destructive interference would have little impact on the constraint the LHC data places upon the multiscale model, and progressively greater impact on the viability of the $N=2$ versions of the isotriplet, TCSM low-scale, and original one-family models.  In the variant one-family model, we see that the contribution from the top loop would, as discussed above, dominate at lower $m_P$, cancel the techniquark loop contribution at $m_P \approx 300$ GeV (so that the expected cross-section would vanish), and then diminish in size for larger $m_P$.

We have also explored the impact of top loop contributions with $\epsilon_t = 0.5$ on the $N_{TC}=4$ versions of the models, where the value of $R^{loops}$ would be smaller by a factor of two.  We find that destructive interference from top loops would leave the LHC data's exclusion of technipions intact across the range 160 GeV $< m_P < 2m_t$ in the multiscale model, would bring the upper end of the excluded range down to 325 GeV (300 GeV, 250 GeV) in the isotriplet (TCSM low-scale, variant one-family) model from the value of $2m_t$ shown in Figure \ref{fig:XSBRggHtautauATLASTau-log}, and bring the upper range of the excluded range down to about 250 GeV from the previous 325 GeV (per Figure \ref{fig:XSBRggHtautauATLASTau-log}) in the original one-family model.  The impact on models with even larger values of $N_{TC}$ would be proportionately smaller.

Finally, we note that if data were available for di-photon final states in the applicable mass range, it would be possible to discern the impact of destructive interference between top and technifermion loops on the data's ability to constrain the models.  In this case, one would need to include effects of top-quark loops both on technipion production from gluon fusion and also on technipion decay to two photons.

\section{Discussion and Conclusions}
\label{axax}

In this paper we have used the LHC limits on the $\gamma\gamma$ \cite{Collaboration:2011ww, CMS-higgs-di-gamma}
 and $\tau^+ \tau^-$ \cite{ATLAS-higgs-di-tau-low, ATLAS-higgs-di-tau-high, CMS-higgs-di-tau} decay modes of a standard model Higgs boson to
 constrain the technipion states predicted in technicolor models with colored technifermions.  As discussed in  \cite{Belyaev:2005ct}, the technipions tend to  produce larger signals in both channels than $h_{SM}$ would, so that this is an effective way of constraining such technicolor models.  Because the technipions are spinless, just like the standard model Higgs boson, the di-photon and di-tau final states resulting from decay of the produced boson would have the same kinematic properties, so there should be no change in the efficiencies and acceptances.  Hence, it is possible to adapt the limits quoted by the collaborations for the Higgs searches very directly to technicolor models with colored technifermions.
 
 We have found that the combined limits on Higgs bosons decaying to di-photon or di-tau final states from the ATLAS and CMS collaborations exclude at 95\% CL  the presence of technipions in the mass range from 110 GeV nearly up to $2 m_t$ for any of the representative models considered here for $N_{TC} \geq 3$.  Even if one takes $N_{TC} = 2$ to make the production rate as small as possible, the multiscale \cite{Lane:1991qh} and isotriplet \cite{Manohar:1990eg} models are excluded up to $2m_t$; the TCSM low-scale  \cite{Lane:1999uh} is excluded for technipion masses up to nearly 300 GeV; and the original \cite{Farhi:1980xs}  and variant  \cite{Casalbuoni:1998fs} one-family model are only marginally consistent with data.   The implication for technicolor model building is clear: models with light technipions and colored technifermions are not allowed by the LHC data, except possibly in a few models with $N_{TC} = 2$.  Model-builders will need to consider scenarios with heavier pseudo Nambu-Goldstone bosons or theories in which the technifermions are color-neutral.

 Moreover, we have also seen that the ATLAS limits on a scalar decaying to $\tau^+\tau^-$ constrain the presence of technipions in the mass range $2 m_t < M_P < $ 450 GeV if  the technipion decays only negligibly to top quarks -- as in models where the top quark's mass is being generated by a topcolor \cite{Hill:1991(a)t} sector instead of by extended technicolor.  The excluded mass range extends to 450 GeV (375 GeV) for a multi scale (isotriplet) technicolor sector for any value of $N_{TC}$ and reaches 375 GeV for a TCSM low-scale technicolor sector with $N_{TC} \geq 3$.  Hence, starting from these technicolor sectors, 
 building a topcolor-assisted technicolor \cite{Hill:1994hp} model would now require ensuring that the technipions have masses above 375 - 450 GeV.  This complements recent  LHC searches for $H \to WW, ZZ$ that exclude the top-Higgs state of TC2 models for masses below 300 GeV if the associated top-pion has a mass of 150 GeV (the lower bound rises to 380 GeV if the top-pion mass is at least 400 GeV) \cite{Chivukula:2011dg}.
 
In principle, there are several ways to construct technicolor models that could reduce the scope of these limits.   As discussed earlier, one possibility is to arrange for the extended technicolor sector to provide a large fraction of the top quark's mass (though it would be necessary to find a new way to evade bounds on FCNC and weak isospin violation).  In this case, gluon fusion through a top-quark loop (as in Figure 1(c)) could provide an alternative production mechanism for the technipions.  If the ETC structure of the model caused the top-quark and techniquark loop amplitudes to interfere constructively, our bounds would be strengthened; but, as illustrated in Figure 5, in a model where the interference was destructive, our limits on the technipion mass could be weakened, at least for small values of $N_{TC}$.  

Another possibility is to build a technicolor model that includes technipions but not colored technifermions\footnote{One example is the ``minimal walking technicolor" model in \cite{Sannino:2004qp} with technifermions in the symmetric tensor representation and $N_{TC} = 2$; various aspects of its collider phenomenology have been predicted, for instance, in \cite{Foadi:2007ue, Belyaev:2008yj}}.  In order for extended technicolor to provide mass to the quarks, color must then be embedded in the ETC group alongside technicolor, and some ETC gauge bosons will carry color charge.  It would be more difficult to use the LHC data discussed here to set broadly-applicable limits on technipions appearing in such models.   The production mechanism contributing most strongly to the rate for the states we studied would not be operative; that is, without colored technifermions, the process illustrated in Figure 1(a) would be absent.  The analogous process with top quarks instead of colored technifermions in the loop (as in Figure 1(c)) could, in principle, contribute, but there will be no loop-derived enhancement by $N_{TC}$ as in the diagram of Figure 1(a).  If the coupling of the top quarks to the technipion were large, that could provide an enhancement to replace the missing $N_{TC}$ factor -- but, as we have seen, the coupling is highly model-dependent.  And, as mentioned earlier, building a model where ETC provides most of the top quark mass (and the top-technipion coupling is large) remains an open challenge, because it is hard to accomplish this without contravening experimental limits on flavor-changing neutral currents  \cite{Dimopoulos:1979es, Eichten:1979ah}  or isospin violation \cite{Chivukula:1988qr}.

A third option would be to base a model around a technicolor sector devoid of technipions, such as the original one-doublet model of \cite{Weinberg:1975gm, Weinberg:1979bn, Susskind:1978ms} or a modern ``next-to-minimal" walking technicolor model with technifermions in the symmetric tensor representation of technicolor and $N_{TC} = 3$ \cite{Sannino:2004qp}.  Of course, these models come with their own complexities and challenges.

This first set of LHC data has excluded a large class of technicolor and topcolor-assisted technicolor models that include colored technifermions -- unless the technipions states can be made relatively heavy or the extended technicolor sector can be arranged to cause interference between top-quark and techniquark loops.  Model builders will need to either identify specific technicolor theories able to withstand the limits discussed here, while generating the top quark mass without excessive weak isospin violation or FCNC,  or else seek new directions for a dynamical explanation of the origin of mass.   Finally, we would like to stress that additional LHC data that gives greater sensitivity to new scalars decaying to $\tau^+\tau^-$ or that addresses scalars with masses over 145 GeV decaying to $\gamma\gamma$ could quickly probe models down to the minimum number of technicolors and up to higher technipion masses.

\bigskip

\begin{acknowledgments}
The work of RSC and EHS was supported, in part, by the US National Science Foundation under grant PHY-0854889. JR is supported by the China Scholarship Council.  PI was supported by the Thailand Development and Promotion of Science and Technology Talents Project
(DPST). 
\end{acknowledgments}


\end{document}